\documentclass[12pt]{iopart}

\usepackage[T1]{fontenc}
\usepackage{subfigure}

\usepackage{graphicx}
\usepackage{mathrsfs}
\usepackage[british]{babel}

\expandafter\let\csname equation*\endcsname\relax 
\expandafter\let\csname endequation*\endcsname\relax 
\usepackage{amsbsy,amssymb,amsmath,bm}
\usepackage{graphicx,color,epsfig,rotate}
\usepackage{braket}
\usepackage{tikz}
\usetikzlibrary{intersections,decorations.pathmorphing,decorations.markings,decorations.pathreplacing,decorations.shapes}
\usepackage{booktabs}

\usepackage{color}
\newcommand{\rev}[1]{#1}

\newcommand{\n}[1]{\left| #1 \right|}
\newcommand{\setc}[2]{\left\{#1\; :\; #2 \right\}}
\renewcommand{\v}[1]{\boldsymbol{#1}}
\newcommand{\R}{\mathbb{R}}

\newcommand{\Z}{\mathbb{Z}}

\newcommand{\C}{\mathbb{C}}

\def\Xint#1{\mathchoice
   {\XXint\displaystyle\textstyle{#1}}%
   {\XXint\textstyle\scriptstyle{#1}}%
   {\XXint\scriptstyle\scriptscriptstyle{#1}}%
   {\XXint\scriptscriptstyle\scriptscriptstyle{#1}}%
   \!\int}
\def\XXint#1#2#3{{\setbox0=\hbox{$#1{#2#3}{\int}$}
     \vcenter{\hbox{$#2#3$}}\kern-.5\wd0}}
\def\ddashint{\Xint=}

\def\DDD{\widetilde{\Delta}^{\partial\overline{\partial}}}

\tikzset{
	dot/.style={draw,circle,inner sep=1.5pt,fill=black},
	empty dot/.style={draw,circle,inner sep=1.5pt,fill=white},
	middlearrow/.style={
        decoration={markings,
            mark= at position 0.75 with {\arrow{#1}} ,
        },
        postaction={decorate}
    },
	on each segment/.style={
    decorate,
    decoration={
      show path construction,
      moveto code={},
      lineto code={
        \path [#1]
        (\tikzinputsegmentfirst) -- (\tikzinputsegmentlast);
      },
      curveto code={
        \path [#1] (\tikzinputsegmentfirst)
        .. controls
        (\tikzinputsegmentsupporta) and (\tikzinputsegmentsupportb)
        ..
        (\tikzinputsegmentlast);
      },
      closepath code={
        \path [#1]
        (\tikzinputsegmentfirst) -- (\tikzinputsegmentlast);
      },
    },
  },
}

\catcode`@=11 
\renewcommand\tableofcontents{%
  \section*{\contentsname}%
  \@starttoc{toc}%
}
\catcode`@=12

\usepackage{multirow}

\makeatletter
\@addtoreset{footnote}{section}
\makeatother

\begin{document}

\title[Entanglement Entropy in Excited States of the Quantum Lifshitz Model]{Entanglement Entropy in Excited States of the Quantum Lifshitz Model}

\author{Daniel E. Parker$^1$, Romain Vasseur$^{1,2}$, and Joel E. Moore$^{1,2}$}
\address{${}^1$Department of Physics, University of California, Berkeley, California 94720, USA}
	\address{${}^2$Materials Sciences Division, Lawrence Berkeley National Laboratory, Berkeley, California 94720, USA}

\eads{\mailto{daniel\_parker@berkeley.edu},
	  \mailto{rvasseur@berkeley.edu}, 
      \mailto{jemoore@berkeley.edu}}

\date{\today}

\begin{abstract} 
We investigate the entanglement properties of an infinite class of excited states in the quantum Lifshitz model (QLM). The presence of a conformal quantum critical point in the QLM makes it unusually tractable for a model above one spatial dimension, enabling the ground state entanglement entropy for an arbitrary domain to be expressed in terms of geometrical and topological quantities. Here we extend this result to excited states and find that the entanglement can be naturally written in terms of quantities which we dub ``entanglement propagator amplitudes'' (EPAs). EPAs are geometrical probabilities that we explicitly calculate and interpret.  A comparison of lattice and continuum results demonstrates that EPAs are universal. This work shows that the QLM is an example of a 2+1d field theory where the universal behavior of excited-state entanglement may be computed analytically.

\end{abstract}
 

\newpage

\section{Introduction}

The tools of quantum information theory have proven increasingly useful in recent years to understand and characterize many-body quantum systems~\cite{RevModPhys.80.517,Laflorencie2016}. A prime example is the use of entanglement entropy to determine if a groundstate wave function is critical or topologically ordered. Remarkably, the entanglement structure of many-body quantum groundstates of systems with local interactions is quite special, and is quite different from that of random states. This result is known as the ``area law''~\cite{Srednicki1993,1742-5468-2007-08-P08024}: in the ground state of a model, the entanglement between two subsystems is usually proportional to the surface area between them, whereas in highly excited states or for a random state in the Hilbert space it usually becomes proportional to the volume~\cite{Srednicki1993,Page:1993aa,Eisert2010}. These are robust results that are not model-dependent, and they also have crucial implications for numerical techniques as they explain the success of variational matrix-product state algorithms such as the density matrix renormalization group~\cite{white92,PhysRevLett.91.147902,whitetdmrg,schollwoeck}. 

	Defining the entanglement entropy requires an ultraviolet cutoff (lattice spacing) and is therefore not completely universal, but the entanglement often contains universal pieces that are characteristic of a model independently of how it is written. Many examples of this are now known. In a 1+1d conformal field theory, for instance, the entanglement entropy scales logarithmically with the size of the entanglement interval times a coefficient proportional to the central charge of the theory~\cite{Holzhey1994,PhysRevLett.90.227902,Calabrese:2004hl,Calabrese:2009dx}. In topological systems, the number of topological degrees of freedom is reflected in a constant term in the entanglement~\cite{PhysRevLett.96.110404,PhysRevLett.96.110405}. Even strongly disordered systems can sometimes show universal logarithmic corrections to the area law in one dimension~\cite{PhysRevLett.93.260602,1751-8121-42-50-504010}.  One can therefore read off many physically interesting quantities and understand a good deal of the structure of the theory by knowing the entanglement. 

Although most results are restricted to low-energy equilibrium settings, the scaling of entanglement in highly excited states or after a global quantum quench has attracted a lot of attention in the past few years~\cite{PhysRevLett.96.136801,1742-5468-2005-04-P04010,Calabrese2016}. In particular, it was realized that entanglement is a key tool to understand how isolated many-body quantum systems self-thermalize (or fail to do so) under their own unitary dynamics~\cite{2014arXiv1404.0686N,Altman:2015aa,1742-5468-2016-6-064010}. For example, in many-body localized systems, the area law is preserved even in highly excited states, a fact closely related to the non-ergodic behavior of those systems~\cite{BauerNayak,PhysRevLett.111.127201}. \rev{Results on one dimensional systems also indicate that the entanglement entropy of low-lying excited states may have universal contributions~\cite{Alcaraz:2011gx,Alba:2009em,bucciantini2014quantum,berganza2012entanglement,he2014quantum}}.

Unfortunately the promise of using entanglement to understand condensed matter systems is mitigated by the difficulty of computing it. The best ``playground'' for entanglement is 1+1 dimensional conformal field theories, where a field theory replica technique pioneered by Cardy and Calabrese~\cite{Calabrese:2004hl,Calabrese:2009dx} provides a practical computational framework for the ground states, at finite temperature~\cite{Calabrese:2004hl}, and for dynamics~\cite{Calabrese2016}. These ideas have been extended to free field theories in higher dimensions~\cite{Casini:2009cz} and to specialized interacting theories using the AdS/CFT correspondence~\cite{PhysRevLett.96.181602,1751-8121-42-50-504008}. There are especially few results concerning the entanglement of excited states in dimensions larger than one, both in highly excited states where entanglement should be a prime tool to study questions related to thermalization, and in low-energy excited states that may show interesting universal contributions to the entanglement.

As a step towards this goal, this paper will describe the entanglement properties in excited states of the 2+1d quantum Lifshitz Model (QLM). \rev{The QLM} arose as a continuum limit of the venerable quantum dimer model~\cite{Rokhsar:1988zz,PhysRevLett.86.1881,PhysRevB.65.024504}, which began as model for the resonating-valence-bond theory of high temperature superconductivity with a quantum critical point known as the RK point~\cite{Rokhsar:1988zz}. (See~\cite{Moessner:2008aa} for a review). The QLM also provides an example of a deconfining critical point~\cite{Vishwanath:aa,fradkin2004bipartite}. The quantum Lifshitz model is a continuum limit of the dimer model at the RK point via a height function~\cite{Henley:1996iw,Kenyon2001}. It is an example of a conformal quantum critical point, whose wave\textit{functional} has 2d conformal invariance, so that techniques of conformal field theory (CFT) are applicable. This special property simplifies the entanglement of the theory and makes analytic progress possible. It was argued in~\cite{Fradkin:2006hs} (and subsequently confirmed in~\cite{Hsu:2008fc,Stephan2009,Fradkin:2009hp,Oshikawa2010, Zaletel:2011dz,Zhou2016a} with slight corrections) that the entanglement entropy in arbitrary geometries may be expressed in terms of the free energy $F$ of the associated CFT. One can hope that the understanding of entanglement in excited states of the QLM will inform excited-state studies in standard ($z=1$) quantum critical points with conformal invariance in $d+1$ dimensions, similarly to how the corner contributions to entanglement in the QLM can be compared to the corner contributions in the $z=1$ case, where much has been learned even in the absence of a closed form~\cite{witczak-krempa}.

Let us quickly recall the results of \cite{Fradkin:2006hs}. Suppose the QLM is defined on a 2d spatial manifold $M$, with a partition $A \cup B = M$. Fradkin and one of us showed that the entanglement entropy can be expressed in terms of free energies of the 2d CFT describing the ground state wavefunctional (in most physically relevant cases, a free boson)
\begin{equation}
	S_A = F_A + F_B - F_{A \cup B} + {\cal O}(1)_\text{topological}.
	\label{eq:S_QLM_GS_I}
\end{equation}
The scaling of the entanglement entropy then follows from a standard results of Cardy and Peschel \cite{Cardy:1988ic} that states that if region $A$ is finite \rev{and} simply connected with a smooth boundary, then the free energy scales as 
\begin{equation}
F_A = f_b \n{A} + f_s L - \frac{c \chi}{6} \ln L + {\cal O}(1)
\label{eq:free_energy_scaling}
\end{equation}
where $f_b$ and $f_s$ are non-universal bulk and surface contributions, $\n{A}$ is the area, $c$ is the central charge of the CFT ($c=1$ in our case), $L$ is the perimeter of $A$, and $\chi$ is the Euler characteristic of $A$, $\chi = 2-2g -b$ where $g$ is the genus and $b$ is the number of boundaries. The $\n{A}$ terms cancel in the entropy, yielding a logarithmic correction to the area law determined purely by the geometry and topology:
\begin{equation}
	S_A = 2 f_s L - \frac{c}{6}\Big( \chi_A + \chi_B - \chi_{A\cup B} \Big) \ln L + {\cal O}(1).
	\label{eq:S_QLM_GS_II}
\end{equation}

\rev{Inspired by this concise and general result, this paper explores what may be said about entanglement in excited states of the quantum Lifshitz model. We find that the entanglement of an infinite class of eigenstates is given in terms of geometrical quantities that have a natural probabilistic interpretation, which we dub ``entanglement propagator amplitudes'' (EPAs) for convenience. Intuitively, the three EPAs \eqref{eq:a}--\eqref{eq:d}, can be thought of as the amount of ``entanglement scattering'' from side $A$ to itself, since $B$ to itself, and between $A$ and $B$.
	
	The remainder of this paper is organized as follows. Section 2 describes the QLM in the wavefunctional picture and computes its excited states. Section 3 introduces the EPAs and shows that arbitrary R\'enyi entropies are expressed naturally in terms of EPAs through a combinatorial formula, Equation \eqref{eq:Renyi_entropies_alpha_beta_gamma}. To understand the significance and universality of EPAs, as well as to form an intuitive picture of them, it is necessary to work on the lattice. Section 4 rederives the result \eqref{eq:S_QLM_GS_II} on the lattice via a new method employing the renormalization group and provides numerical checks in tetromino geometries. Section 5 extends this to excited states, defining lattice versions of EPAs. In this setting a physical interpretation is natural and a structure of cancelling divergencs is manifest. Section 6 returns to the continuum to compute EPAs in explicit geometries, presented in Equation \eqref{eq:alpha_beta_gamma_continuum}. A comparison of lattice and continuum results shows that EPAs are numerically equal in both settings --- strong evidence for universality. The appearance of such universal quantities is evidence that studying entanglement of excited states in higher dimensions reveals novel physical quantities of interest. 

}

\section{The Quantum Lifshitz Model}

The quantum Lifshitz model (QLM) is a 2+1d field theory which may be defined on a general 2d spatial manifold $M$. Let $\varphi$ be a compact scalar field (i.e. a field whose target is the circle with radius $2\pi R$) with Dirichlet boundary conditions on $M$. Then define~\cite{Fradkin:2013ww}
\begin{equation}
	H = \frac{1}{2} \int_{M} d^2x \; \left[ \Pi^2 + \kappa^2 \left( \Delta \varphi \right)^2 \right].
	\label{eq:QLM_hamiltonian}
\end{equation}
Here $\Pi(x)$ is the cannonical conjugate momenta, $\Delta$ is the Laplacian on $M$, and $\kappa$ is a real parameter. The value of $\kappa = (8 \pi)^{-1}$ is often used to match correlation functions at the RK point of the quantum dimer model. The QLM (with different choices of $R$ and $\kappa$) is also the continuum limit of various interacting generalizations of the dimer model, as well as the quantum six- and eight-vertex models~\cite{Fradkin:2009hp}.

The QLM is a Gaussian field theory and thus we expect it to be exactly solvable as an infinite collection of harmonic oscillators. It is convenient to use the Schr\"odinger picture to find the \textit{wavefunctionals} $\Psi[\varphi] = \braket{[\varphi]|\Psi}$, which are complex-valued functionals on the space of field configurations, whose eigenfunctionals satisfy $H \Psi[\varphi] = E \Psi[\varphi]$. This treatment expands the ideas of Ardonne \textit{et al.}~\cite{Ardonne:2003gx}, and a pedagogical treatment can be found in Fradkin's textbook~\cite{Fradkin:2013ww}. In this picture, the conjugate momentum becomes a functional derivative $\Pi(x) = -i \delta/\delta \varphi(x)$. We can then take the ``square root'' of the Hamiltonian by defining 
\begin{equation}
	Q^\dagger(x) = \frac{1}{\sqrt{2}}\left( -\frac{\delta}{\delta \varphi(x)} + \kappa \Delta \varphi \right),
	\label{eq:square_root_operator}
\end{equation}
which satisfies $[Q^\dagger(x),Q(y)] = - \kappa \Delta \delta(x-y)$. Then the normal-ordered Hamiltonian becomes $H = \int d^2x \; Q^\dagger(x) Q(x)$.

To find the ground state wavefunctional, note $H$ is positive semi-definite, so any non-trivial eigenfunctional with eigenvalue zero must be the ground state. We thus seek a functional $\Psi_0[\varphi]$ in the kernel of $Q$: $Q(x) \Psi_0[\varphi] = 0$ for all $x$. This is a first-order functional differential equation whose solution is
\begin{equation}
	\Psi_0[\varphi] = \braket{[\varphi]|\Psi_0} = \frac{1}{\sqrt{Z}} e^{-\frac{1}{2} S[\varphi]} \text{ where } Z = \int \mathcal{D}\varphi\;  e^{-S[\varphi]}, \text{ and } S[\varphi] = \int_{M} d^2x \; \kappa \left( \nabla \varphi \right)^2.
	\label{eq:ground_state_wavefunctional}
\end{equation}
This is the action for a compact boson CFT (up to normalization). The QLM is thus an example of a conformal quantum critical point, whose \textit{wavefunctional} has the form of the free 2d boson (even though the 2+1d theory~\eqref{eq:QLM_hamiltonian} has dynamical exponent $z=2$ and is {\it not} conformally invariant), so one may employ the results of 2d conformal field theory when working with the wavefunctional~\cite{Ardonne:2003gx}.

The full spectum of \eqref{eq:QLM_hamiltonian} now follows from properties of the Laplace operator on an arbitrary conected 2d manifold. The collection of all eigenfunctions (or `modes') $\left\{L_\lambda\right\}$ of the Laplace operator on $M$,
\begin{equation}
	\Delta L_\lambda(x) + \lambda L_\lambda(x) = 0,
	\label{eq:eigenfunctions_of_Laplacian}
\end{equation}
have positive eigenvalues $\lambda > 0$ and together form a complete, orthonormal basis for ($L^2$) functions on $M$ \cite{Jost2008}. For each mode $\lambda$,\footnote{Here $\lambda$ or $\mu$ will be used as an index for modes even though finitely many modes may have the same eigenvalue.} define a raising operator 
\begin{equation}
	A^\dagger_\lambda = \frac{1}{\sqrt{\kappa \lambda}} \int_{M} d^2x \; L_\lambda(x) Q^\dagger(x).
	\label{eq:raising_operator}
\end{equation}
By using the above commutation relation for the $Q$'s, one may show that that, if $\lambda$ and $\mu$ are two eigenvalues, $[A_\lambda, A^\dagger_\mu] = \delta_{\lambda \mu}$, $[A_\lambda^\dagger, A_\mu^\dagger] = [A_\lambda, A_\mu] = 0$, and $[H, A_\lambda^\dagger] = \kappa \lambda A_\lambda^\dagger$, and thus a general energy eigenstate is indexed by the number of quanta $n_\lambda$ in each mode $\lambda$ of the Laplace operator. We may then rewrite the Hamiltonian for the wavefunctionals as
\begin{equation}
	H = \sum_{\lambda} \kappa \lambda A^\dagger_\lambda A_\lambda. 
	\label{eq:hamiltonian_modes}
\end{equation}
It is now clear that the wavefunctional of arbitrary excited states is directly analogous to the quantum harmonic oscillator:
\begin{equation}
	\braket{[\varphi]|(n_{\lambda_1}, n_{\lambda_2},\cdots)} = \prod_{\lambda} \left( Z \; 2^{n_\lambda} n_\lambda! \right)^{-1/2} H_{n_\lambda}\left( \sqrt{\kappa \lambda} \varphi^\lambda \right) e^{-\frac{1}{2} S[\varphi]}.
	\label{eq:excited_wavefunctional}
\end{equation}
where $H_n$ is the $n$th Hermite polynomial and $\varphi^\lambda = \int_{M} d^2x \; L_\lambda(x) \varphi(x)$.

\section{R\'enyi Entropy in Excited States}
\label{sec:continuum_excited_states}

Now equipped with the excited states of the model, we can look to understand their entanglement. \rev{Using a generalization of Wick's theorem, this section derives a formula, Equation \eqref{eq:Renyi_entropies_alpha_beta_gamma}, for arbitrary R\'enyi entropies in excited states.}

For definiteness, let us work in the state with one quantum in the $\lambda$th mode, whose wavefunctional is therefore
\begin{equation}
	\Psi_\lambda[\varphi] = 	\braket{[\varphi]|\Psi_\lambda} = Z^{-1/2} \sqrt{2 \kappa \lambda }\;  \varphi^\lambda e^{-\frac{1}{2} S[\varphi]}. 
	\label{eq:one_mode_wavefunctional}
\end{equation}
This is an artificial restriction to make the calculations easier and the exposition clearer; the method below works for any wavefunctional. However, the combinatorics is prohibitively difficult to make exact statements for arbitrary states.

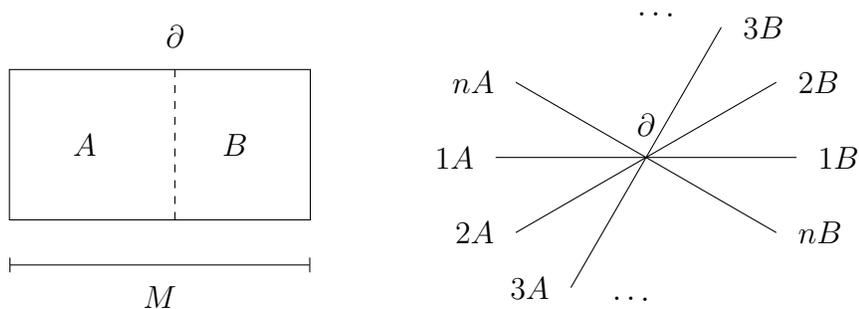
\begin{figure}[h]
	\center
		\begin{tikzpicture}[scale=2]
		\draw (0,0) rectangle (2,1);
		\draw[dashed] (1.1,0) -- (1.1,1) node[label=above:$\partial$] {};
		\node at (0.5,0.5) {$A$};
		\node at (1.5,0.5) {$B$};
		\draw[|-|] (0,-0.3) -- (2,-0.3) node[midway,label=below:$M$] {};
	\end{tikzpicture}
	\hspace{3em}
\begin{tikzpicture}
	\draw (0:-2) node[label=left:$1A$] {} -- (0:2) node[label=right:$1B$] {};
	\draw (30:-2) node[label=left:$2A$] {} -- (30:2) node[label=right:$2B$] {};
	\draw (-30:-2) node[label=left:$nA$] {} -- (-30:2) node[label=right:$nB$] {};
	\draw (60:-2) node[label=left:$3A$] {} -- (60:2) node[label=right:$3B$] {};
	\node[rotate=-3] at (85:1.9) {$\cdots$};
	\node[rotate=-3] at (85:-1.9) {$\cdots$};
	\node[label=above:$\partial$] at (0,0) {};
\end{tikzpicture}
\caption{(Left) An example of the geometry of ${M}$. (Right) A side-on schematic of the $n$-sheeted surface $\mathcal{S}_n$.}
\label{fig:n-sheet-geometry_continuum}
\end{figure}

To set notation, let us say that ${M} = A \cup B$ and let $\partial$ be the border between $A$ and $B$ --- see Figure \ref{fig:n-sheet-geometry_continuum}. Then we may write a field-configuration as $\varphi = \varphi^A \oplus \varphi^B$, the sum of fields with support on $A$ and $B$ respectively. The density matrix $\rho = \ket{\Psi_\lambda} \bra{\Psi_\lambda}$ has a partial trace with matrix elements
\rev{
\begin{equation}
	\braket{[\varphi^1_A] | \rho_A | [\varphi^2_A]} = \int \mathcal{D} \varphi_B \braket{[\varphi_A^1\oplus \varphi_B]|\rho|[\varphi_A^2 \oplus \varphi_B]}.
	\label{eq:partial_trace}
\end{equation}
}
The path integral is over all possible field configurations on the $B$ side of the manifold. \rev{Each field (e.g. $\varphi^2_A$) lives on one sheet on one side (sheet two on side $A$), as shown in Figure \ref{fig:n-sheet-geometry_continuum}.} To find an arbitrary R\'enyi entropy $S^{(n)}_A = \frac{1}{1-n} \ln \Tr \rho_A^n$ we must compute 
\rev{
\begin{equation}
	\Tr \rho_A^n 
	= \int \prod_{a \in \Z_n} \mathcal{D}\varphi^a_A \mathcal{D} \varphi^a_B  \braket{[\varphi_A^a \oplus \varphi^a_B]|\rho|[\varphi^{a+1}_A \oplus \varphi^a_B]} 
	= \int\limits_{\mathcal{S}_n} \mathcal{D} \varphi^a \braket{[\varphi_A^a \oplus \varphi^a_B]|\rho|[\varphi^{a+1}_A \oplus \varphi^a_B]}, 
	\label{eq:arb_Renyi_path_integral}
\end{equation}
}
where $\int_{\mathcal{S}_n}$ is the path integral over all field configurations on ${M}$ with the constraint that each of the \rev{$\varphi^a$'s} are equal on $\partial$~\cite{Fradkin:2006hs}. This comes from the fact that a discontinuity of the fields across  the boundary gives an  infinite contribution to the action, so the fields must agree at the boundary~\cite{Fradkin:2006hs}. Henceforth, the product over $a \in \Z_n$ will be implicit. 
One may consider this as a path integral over an $n$-sheeted surface $\mathcal{S}_n$ made of $n$ copies of $M$, which are pinned together along $\partial$. This is \textit{quite} different than the Riemann surfaces that often show up in 1+1d CFT calculations of entanglement~\cite{Calabrese:2004hl}; the surface $\mathcal{S}_n$ is not a manifold (since it is not homeomorphic to a neighborhood of $\R^2$ near $\partial$) and may have the cut anywhere, not just along a spatial axis. Unfortunately, this precludes the possibility of a conformal mapping to a standard geometry.

Define the \textit{surface expectation} to be
\rev{
\begin{equation}
	\braket{\mathcal{O}(x_1) \mathcal{O}(x_2)\cdots }_{\mathcal{S}_n} = Z^{-n} \int_{\mathcal{S}_n} \mathcal{D} \varphi_a \; \Big( \mathcal{O}(x_1) \mathcal{O}(x_2)\cdots \Big) e^{-\sum_{a=1}^n S[\varphi^a]}.
	\label{eq:surface_expectation}
\end{equation}
}
The problem of finding the R\'enyi entropies can therefore be rephrased as computing the expectation value
\rev{
\begin{equation}
	\Tr \rho_A^n = \left( 2\kappa \lambda \right)^n \Braket{ \varphi^{11}_\lambda \varphi^{21}_\lambda\varphi^{22}_\lambda \cdots \varphi^{nn}_\lambda \varphi^{1n}_\lambda}_{\mathcal{S}_n}
	\label{eq:Renyi_as_surface_expectation}
\end{equation}
}
where
\rev{
\begin{equation}
	\varphi^{ab} = \begin{cases}
	\varphi^a(x) & x \in A\\
	\varphi^b(x) & x \in B
	\end{cases}
	\quad
	\text{ and }
	\quad
	\varphi_\lambda^{ab} = \int_{M} d^2x \; L_\lambda(x) \varphi^{ab}(x).
	\label{eq:split_field_configuration}
\end{equation}
}

\subsection{The Sides and the Boundary}

The task is now to evaluate the correlation functions in \eqref{eq:Renyi_as_surface_expectation}. By linearity of expectation,
\rev{
\begin{equation}
	\Tr \rho_A^n = \left( 2\kappa \lambda \right)^n  \int_{M} \prod_{i=1}^{2n} d^2 x_i\; L_\lambda(x_i) \braket{\varphi^{11}(x_1) \varphi^{21}(x_2) \cdots}_{\mathcal{S}_n},
	\label{eq:spatial_correlations_only}
\end{equation}
}
so all that is necessary is to evaluate spatial $2n$-point functions on $\mathcal{S}_n$. \ref{app:wick_thm} derives Wick's theorem --- suitably modified for this geometry --- by applying a series of coordinate transformations first developed in the Appendix of~\cite{Zaletel:2011dz} and simplified in~\cite{Zhou2016a}. If the $\varphi_{a_i}(x_i)$'s are fields on sheets $a_i$, then the result is
\rev{
\begin{equation}
	\begin{aligned}
	&\frac{\braket{\varphi^{a_1}(x_1) \varphi^{a_2}(x_2) \cdots \varphi^{a_{2n}}(x_{2n})}_{\mathcal{S}_n}}{\Tr \rho_A^n(\text{GS})}  \\
		& =  \left( 2\kappa \right)^{-n}\sum_{\text{``pairs"}} 
		\prod_{i=1}^n
		\left( \delta^{a_{c_i} a_{d_i}}  G^D(x_{c_i}, x_{d_i})  + \frac{1}{n} \left[ G^{M}(x_{c_i},x_{d_i}) - G^{D}(x_{c_i},x_{d_i}) \right] \right) + \text{Har}
	\end{aligned}
	\label{eq:wick_modified}
\end{equation}
}
where
\begin{equation}
	\text{``pairs''} = \setc{C = (c_1,c_2,\dots,c_n), D = (d_1, d_2,\dots,d_n) \subset \Z_{2n}}{C\cup D = \Z_{2n}},
	\label{eq:pairs_as_set}
\end{equation}
i.e. all possible partitions of $2n$ elements into pairs. The notation $G^{M}$ (\text{resp}. $A$, $B$) is the Green's function for the Laplacian on ${M}$ (\rev{\textit{resp}.} $A$, $B$) with Dirichlet boundary conditions,
and \begin{equation}
	G^D(x,y) = \begin{cases}
		G^A(x,y) & x,y \in A\\
		G^B(x,y) & x,y \in B\\
		0 & \text{else.}
	\end{cases}
	\label{eq:dirichlet_Greens_operator}
\end{equation}
``Har'' means this is only true up to an arbitrary harmonic function; this is irrelevant because we are always integrating against eigenfunctions of the Laplacian, so we can remove it with an integration by parts. Lastly, the \rev{normalization factor} is exactly the ground state R\'enyi entropy
\begin{equation}
	\Tr \rho_A^n(\text{GS}) = \mathcal{T} \left( \frac{Z_A Z_B}{Z_M} \right)^{n-1},
	\label{eq:ground_state_Renyi}
\end{equation}
a familar result from, e.g. \cite{Fradkin:2006hs,Zaletel:2011dz,Zhou2016a}, where \rev{$Z_D = \left(\det \Delta_D \right)^{-1/2}$ is the partition function on the domain $D$~\cite{Fradkin:2006hs}} and   $\mathcal{T}$ represents the ${\cal O}(1)$ topological terms. Our generalized Wick's Theorem is therefore the normal Wick's theorem on each side $A$ and $B$ of each sheet, plus a contribution $G^M-G^D$ which communicates between sides and between sheets, but is suppressed by $1/n$.

\subsection{Entanglement Propagator Amplitudes}

Combining \eqref{eq:spatial_correlations_only} and \eqref{eq:wick_modified} then tells us
\begin{equation}
	\begin{aligned}
	&\frac{\Tr \rho_A^n(\Psi_\lambda)}{ \Tr \rho_A^n(\text{GS})}=  \sum_{\text{``pairs"}} \prod_{i=1}^n \lambda \int d^2 x_{c_i} d^2 x_{d_i} L_\lambda(x_i) L_\lambda(y_i)\\
	&\hspace{3em} \times \left( \delta_{a_{c_i} a_{d_i}}(x_{c_i}x_{d_i})  G^D(x_{c_i}, x_{d_i}) + \frac{1}{n} \left[ G^{M}(x_{c_i},x_{d_i}) - G^{D}(x_{c_i},x_{d_i}) \right] \right) ,
	\end{aligned}
	\label{eq:renyi_from_wick}
\end{equation}
where $\delta_{a_{c_i}a_{d_i}}(x_{c_i},x_{d_i})$ is unity only if the fields are on the same sheet. For instance, for the pair of fields $\varphi^{11}(x_1)\varphi^{21}(x_2)$, it is unity only when $x_1, x_2 \in B$. Equation
\eqref{eq:renyi_from_wick} is a polynomial in three quantities, which we dub the ``Entanglement Propagator Amplitudes'' (EPAs),
\begin{subequations}
\begin{align}
	\label{eq:a}
	\alpha \ &=\ \lambda \int_{M} d^2 x \; d^2 y\; L_\lambda(x) G^A(x,y) L_\lambda(x),\\ 
	\label{eq:b}
	\beta \ &=\ \lambda \int_{M} d^2 x \; d^2 y\; L_\lambda(x) G^B(x,y) L_\lambda(x),\\
	\label{eq:d}
	\gamma \ &=\ \lambda \int_{M} d^2 x \; d^2 y\; L_\lambda(x) \left[ G^{M}(x,y) - G^D(x,y) \right] L_\lambda(x).
\end{align}
\end{subequations}
Here $\alpha$ and $\beta$ are intra-sheet terms that have support on $A$ or $B$ respectively. On the other hand, $\gamma$ is an inter-sheet term, connecting $A$ to $B$. These are dimensionless and should be interpreted as probabilities since $0 \le \alpha, \beta, \gamma \le 1$ and $\alpha + \beta + \gamma = 1$. 

Applying a dash of combinatorics,
\begin{equation}
\frac{\Tr \rho_A^n(\Psi_\lambda)}{ \Tr \rho_A^n(\text{GS})}=
	\sum_{\substack{\text{``pairs"}\\C,D}} \prod_{i=1}^n
	\left.	\frac{\gamma}{n} + \begin{cases}
		\alpha & \text{if $c_i + 1 \equiv  d_i  \pmod n$ and $c_i$ even}\\
		\beta & \text{if $c_i + 1 \equiv  d_i  \pmod n$ and $c_i$ odd}\\
		0 & \text{otherwise}
	\end{cases}
\right\}.
\label{eq:Renyi_entropies_alpha_beta_gamma}
\end{equation}
This allows us to compute arbitrary R\'enyi entropies for excited states with one quantum in an arbitrary mode. Such a restriction is artificial; the R\'enyi entropy of an excited state with multiple quanta could be computed with more complex combinatorics and three EPAs for each pair of modes involved. For example if we had two modes $\lambda$ and $\mu$, then the R\'enyi entropies would involve quantities such as
\begin{equation}
	\alpha_{\lambda\mu} \ =\ \sqrt{\lambda \mu} \int_{M} d^2 x \; d^2 y\; L_\lambda(x) G^A(x,y) L_\mu(x).\\ 
	\label{eq:alpha_lambda_mu}	
\end{equation}

\rev{Equation \eqref{eq:Renyi_entropies_alpha_beta_gamma} is a main result of this work. To leverage \eqref{eq:Renyi_entropies_alpha_beta_gamma} to compute entanglement entropy, we must understand both how to compute the EPAs and the combinatorics of how they are combined. This section comments on the combinatorics briefly and we will return to the interpretation and evaluation of the EPAs presently, with a physical interpretation at the end of Section \ref{sec:lattice_excited}.}

To find the entanglement entropy, one must analytically continue \eqref{eq:Renyi_entropies_alpha_beta_gamma} to arbitrary $n$, in practice by finding a closed form equation for it \cite{Calabrese:2004hl}, a non-trivial problem. Explicitly, the first several traces are 
\begin{subequations}
\begin{align}
	\frac{\Tr \rho_A^1(\Psi_\lambda)}{\Tr \rho_A^1(\text{GS})} &= 1, \\
	\frac{\Tr \rho_A^2(\Psi_\lambda)}{\Tr \rho_A^2(\text{GS})} &= a^2+b^2+g^2,\\
	\frac{\Tr \rho_A^3(\Psi_\lambda)}{\Tr \rho_A^3(\text{GS})} &= a^3+3 a b g+3 a g^2+b^3+3 b g^2+4 g^3,\\
	\frac{\Tr \rho_A^4(\Psi_\lambda)}{\Tr \rho_A^4(\text{GS})} &= a^4+4 a^2 b g+8 a^2 g^2+4 a b^2 g + 8 a b g^2\\
	& \quad +20 a g^3+b^4+8 b^2 g^2+20 b g^3+31 g^4,
\end{align}
\label{eq:explicit_renyi_n}
\end{subequations}
where $a = \alpha + d$, $b = \beta + d$, and $g = \gamma/n$. One can translate the problem of computing the R\'enyi entropies into a weighted perfect matching problem. Consider the complete graph on $2n$ vertices, with weights $a$ and $b$ alternating around the outside, and weight $g$ for edges through the middle. The $n$th R\'enyi entropy is the sum of all weighted perfect matchings. (Amusingly, this is equivalent to finding the partition function of a classical dimer problem on a non-planar graph.)

In the special case $\alpha = \beta$, the R\'enyi entropies are polynomials in a single variable $\gamma$. We conjecture that
\begin{equation}
\frac{\Tr \rho_A^n(\Psi_\lambda, \alpha=\beta)}{\Tr \rho_A^n(\text{GS})} 
\ =\ 2^{n-1}\left( 
1+ 
\sum_{k=2}^\infty
\binom{n}{k} (k-1) \; p_k(n) \gamma^k
\right),
\label{eq:all_n_Renyi}
\end{equation}
where
\begin{equation}
	p_k(n) = \sum_{r=0}^{k-2} A_{k-r-2,r} n^r,
	\label{eq:p_k_polynomials}
\end{equation}
wherein the coefficients $A_{i,j}$ are the diagonals of OEIS sequence A112486~\cite{oeis} and may be recursively defined by 
\begin{equation}
	A_{i,j} = \begin{cases}
		0 & j = -1 \text{ or } i < j\\
		1 & i=j=0\\
		(i+j) A_{i-1,j} + (i+j-1) a_{i-1,j-1} & \text{otherwise.}
	\end{cases}
	\label{eq:A_sequences}
\end{equation}
We have confirmed Equation \eqref{eq:all_n_Renyi} explicitly through $n=11$ via computer.

To find the entanglement in general it would be convenient if one of $a$, $b$ or $g$ could be treated as a small expansion parameter. The behavior of \eqref{eq:Renyi_entropies_alpha_beta_gamma} to lowest order in $g = \gamma/n$ is particularly simple: 
\begin{equation}
\frac{\Tr \rho_A^n(\Psi_\lambda)}{ \Tr \rho_A^n(\text{GS})} =   a^n + b^n + {\cal O}\left( \gamma \right),
	\label{eq:alpha_beta_only}
\end{equation}
It is therefore quite tempting to hope that $\gamma$ vanishes, which would make \eqref{eq:all_n_Renyi} trivial. Superficially, it \textit{looks} like $\gamma$ might vanish; we may integrate by parts to get $\Delta G^{M} - \Delta G^D$, which would give a delta function on $M$ minus delta functions on both $A$ and $B$. It would appear that the inter-sheet EPA therefore vanishes, since the integral kernel is identically zero except on the co-dimension one boundary $\partial$. In this case, $\alpha + \beta =1$ and $\alpha$ would be the integral of the wavefunction over $A$, i.e. the probability the excitation is on the $A$ side, and similarly for $B$. The analytic continuation of \eqref{eq:alpha_beta_only} would be trivial, and the entanglement entropy would be that of the ground state plus the entropy of an effective qubit for each quanta with density matrix $\rho_A^\text{eff} =  \operatorname{diag}(\alpha, \beta)$. However we will demonstrate below that, contrary to this naive expectation, $\gamma$ is generically non-zero and is not small in any common physical limit. The reasoning of the above paragraph therefore \textit{does not apply}, though important aspects of it survive, and it is a good first start for intuition.

To see why $\gamma$ is non-zero and what its interpretation is in the continuum involves regularization and subtle arguments. Therefore, we will explore a explicit lattice regularization of the problem, providing a point of view complementary to the field theory picture, but without divergences. Section \ref{sec:lattice_GS} will reproduce the entanglement entropy in the ground state from the lattice, and Section \ref{sec:lattice_excited} will examine lattice excited states, showing $\gamma$ is generically non-zero. With this analysis in hand, Section \ref{sec:continuum_redux} will reveal how this is realized in the continuum. We will conclude that the EPAs $\alpha$, $\beta$, and $\gamma$ are universal and completely geometric quantities which fully determine the entanglement entropy of excited states of the theory. For states with one quantum, it therefore follows that the difference in entanglement with the ground state is $\mathcal{O}(1)$ and universal.

\section{Building the Quantum Drum from Scaffolding}

\rev{To understand and compute the EPAs on the lattice, it is first necessary to reproduce the ground state results in a lattice setting. To that end, this section re-derives \eqref{eq:S_QLM_GS_II} with a lattice regularization and confirms it numerically.}

\label{sec:lattice_GS}

Let us make the following assumptions:
\begin{enumerate}
	\item The only difference in the entanglement between the compact and non-compact theories is an $\mathcal{O}(1)$ topological term. We saw in Section \ref{sec:lattice_GS} that this holds for the ground state and \ref{app:wick_thm} suggests it holds for excited states as well. Therefore we will examine the non-compact lattice theory.
	\item Regularizing the original Hamiltonian \eqref{eq:QLM_hamiltonian}, is the same as regularizing the emergent free-boson Lagrangian \eqref{eq:ground_state_wavefunctional}. 
\end{enumerate}

\begin{figure}[h]

	\begin{tikzpicture}[scale=2]
		\draw (0,0) rectangle (2,1);
		\draw[dashed] (0.9,0) -- (0.9,1);
		\foreach \i in {0,1,...,5}{
			\foreach \j in {0,1,...,10}{
				\node[dot] at (0.2*\j,0.2*\i) {};
			}
		}

		\draw[|-|] (-0.25,0) -- (-0.25,1) node[midway,label=left:$L$] {};
		\draw[|-|] (0,-0.25) -- (0.9,-0.25) node[midway,label=below:$\ell$] {};
		\draw[|-|] (0,-0.6) -- (2,-0.6) node[midway,label=below:$K$] {};
		
		\draw[|-|] (0,1.2) -- (0.9,1.2) node[midway,label=above:$A$] {};
		\draw[|-|] (0.9,1.2) -- (2,1.2) node[midway,label=above:$B$] {};

		\draw[|-|] (0.9,1.4) -- (1.1,1.4) node[midway,label=above:$\overline{\partial}$] {};
		\draw[|-|] (0.7,1.4) -- (0.9,1.4) node[midway,label=above:$\partial$] {};

	\end{tikzpicture}
	\hspace{3em}
	\begin{tikzpicture}
		\foreach \yA in {-1,1}{
			\draw (-4,\yA) -- (0,\yA);
			\foreach \xA in {-4,-3.2,...,0.2}{
				\node[dot] at (\xA,\yA) {};
			}
		}
		\foreach \yB in {-2,0,2}{
			\draw (0.8,\yB) -- (4.8,\yB);
			\foreach \xB in {0.8,1.6,...,4.9}{
				\node[dot] at (\xB,\yB) {};
			}
		}
		\foreach \y in {-3,-1,1}{
			\draw[dashed] (0,\y) -- (0.8,\y+1);
			\draw[dashed] (0.8,\y+1) -- (0,\y+2);
		}
		\node[label=above:$1A$] at (-2,-1) {};
		\node[label=above:$2A$] at (-2,1) {};

		\node[label=above:$2B$] at (2.8,2) {};
		\node[label=above:$1B$] at (2.8,0) {};
		\node[label=above:$nB$] at (2.8,-2) {};
		
		\node[label=above:$\ell$] at (0,1) {};
		\node[label=above:$\ell+1$] at (1,0.2) {};
	\end{tikzpicture}
	\caption{(Left) The lattice geometry we adopt. The column of $A$ next to the boundary is denoted $\partial$ and the column of $B$ next to the boundary is $\overline{\partial}$. (Right) A side view of the discretized $n$-sheeted surface $\mathcal{S}_n$. The dashed lines represent half-strength bonds between the $A$ and $B$ sides.}
	\label{fig:lattice_geometry}
\end{figure}
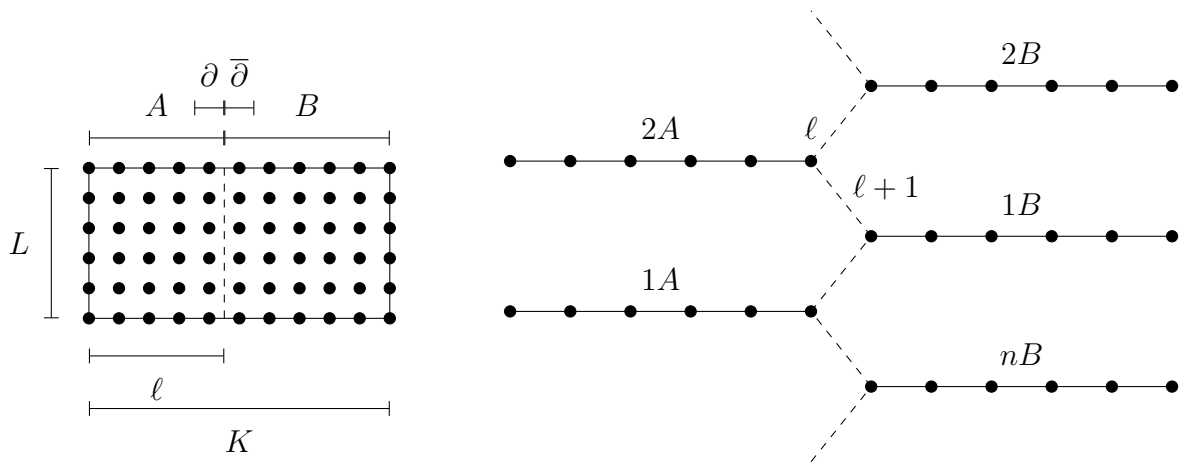

We will work on the square lattice of length $a$. To make calculations concrete, we adopt the explicit geometry of Figure \ref{fig:lattice_geometry}: a rectangle of size $K \times L$ where the $A$ side is $\ell \times L$. Though a specific geometry has been adopted, the Riemann mapping theorem implies these results hold for any geometry where $A$ and $B$ are simply connected. With a variable $\varphi_{pq} \in \R$ at each lattice site, the action becomes (factors of $a$ cancel out)
\begin{equation}
	\widetilde{S}[\varphi_{pq}] = \kappa \sum_{p=1}^K \sum_{q=1}^L \Big[4\varphi_{p,q} - \varphi_{p,q} \left( \varphi_{p+1,q} + \varphi_{p-1,q} + \varphi_{p,q+1} + \varphi_{p,q+1} \right) \Big].
	\label{eq:discrete_action}
\end{equation}
Tilde's will be used to denote the lattice versions of quantities (except for fields) for easy comparison with the continuum equations. Dirichlet boundary conditions are imposed by setting ``out-of-bounds'' variables such as $\varphi_{0,3}$ to zero in the action.

\begin{figure}
	\center
	\begin{tikzpicture}
		\foreach \yA in {-2,0,2}{
			\draw (-4,\yA) -- (0,\yA);
			\foreach \xA in {-4,-3.2,...,0.2}{
				\node[dot] at (\xA,\yA) {};
			}
		}
		\foreach \yB in {-2,0,2}{
			\draw (0.8,\yB) -- (4.8,\yB);
			\foreach \xB in {0.8,1.6,...,4.9}{
				\node[dot] at (\xB,\yB) {};
			}
		}
		\draw (0,0) -- (0.8,0) node[midway,label=below:$1$] {};
		\draw[dotted] (0,2) -- (0.8,2) node[midway,label=below:$\cos(2\pi/n)$] {};
		\draw[dotted] (0,-2) -- (0.8,-2) node[midway,label=below:$\cos(2\pi (n-1)/n)$] {};

		\node[label=above:$1A$] at (-2,0) {};
		\node[label=above:$2A$] at (-2,2) {};
		\node[label=above:$nA$] at (-2,-2) {};

		\node[label=above:$2B$] at (2.8,2) {};
		\node[label=above:$1B$] at (2.8,0) {};
		\node[label=above:$nB$] at (2.8,-2) {};
		
		\node[label=above:$\ell$] at (0,2) {};
		\node[label=above:$\ell+1$] at (0.8,2) {};
	\end{tikzpicture}
	\caption{The lattice geometry after the decoupling coordinate transformation. The $A$ and $B$ sides are held together with bonds with strength $0 \le \cos\left( 2\pi k/n \right) \le 1$. After the RG flow, all the dotted lines are removed, i.e. have strength $0$.}
	\label{fig:lattice_geometry_decoupled}
\end{figure}
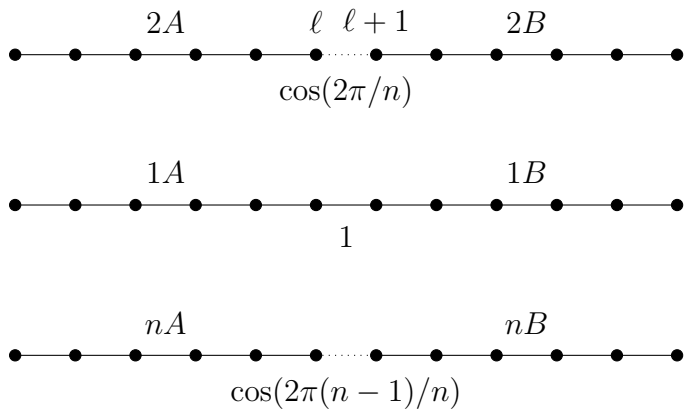

From the definition of the partial trace, the $n$th R\'enyi entropy breaks the ``bonds'' on the boundary between $A$ and $B$ on each sheet and reinstates them with half the weight going to the same sheet and the other half going to the next sheet (see Figure \ref{fig:lattice_geometry}). Therefore the lattice R\'enyi entropies involve the traces\footnote{Here the path integral $\int \mathcal{D}\varphi^c$ is just shorthand notation for $\prod_{pq} \int_\R d\varphi_{pq}$ to show the analogy with the continuum case and not an actual path integral.}
\begin{equation}
	\Tr \widetilde{\rho}_A^n = \frac{1}{\widetilde{Z}_{M}^n} \int \mathcal{D} \varphi^c \exp\left( - \kappa \sum_{c \in \Z_n} \varphi^c_A \widetilde{\Delta}^{AA} \varphi_{A}^{c} +  \varphi^c_B \widetilde{\Delta}^{BB} \varphi_{B}^{c}
	+ \frac{1}{2} \varphi^c_\partial \widetilde{\Delta}^{\partial\overline{\partial}} \varphi_{\overline{\partial}}^{c}
		+ \frac{1}{2} \varphi^c_\partial \widetilde{\Delta}^{\partial\overline{\partial}} \varphi_{\overline{\partial}}^{c+1}
	\right).
	\label{eq:GS_Renyi_entropy_lattice}
\end{equation}
To prevent a plethora of indices, $A$, $B$, $\partial$, $\overline{\partial}$ and $M$ are used as multi-indices to refer to all the lattice sites in those domains, as labelled in Figure \eqref{fig:lattice_geometry}. So $\varphi_{M} \widetilde{\Delta}^{MM} \varphi_{M}$ should be interpreted as $\sum_{(pq),(rs) \in M} \varphi_{pq} \widetilde{\Delta}^{pqrs} \varphi_{rs}$, etc. Here $\widetilde{\Delta}$ is the discrete Laplacian restricted to the appropriate domain. Due to the coupling between sheets, the Gaussian integrals here are non-trivial to evaluate explicitly. However, by applying a Fourier transform we may decouple the sheets. The action may be rewritten in terms of the circulant matrix
\begin{equation}
	T = \begin{bmatrix}
		\frac{1}{2} & \frac{1}{4} & 0 & \cdots & 0 & \frac{1}{4}\\
		\frac{1}{4} & \frac{1}{2} & \frac{1}{4} & \cdots & 0 & 0 \\
		0 & \frac{1}{4} & \frac{1}{2} & \ddots & 0 & 0\\
		\vdots & \vdots & \ddots & \ddots & \ddots &0\\
		0 & 0 & 0 & \ddots & \frac{1}{2} & \frac{1}{4}\\
		\frac{1}{4} & 0 & 0 & \cdots &  \frac{1}{4} & \frac{1}{2}\\
	\end{bmatrix},
	\label{eq:M_n_defn}
\end{equation}
as
\begin{equation}
	\Tr \widetilde{\rho}_A^n = \frac{1}{\widetilde{Z}_{M}^n} \int \mathcal{D} \varphi^c \exp\left( - \kappa \sum_{c\in \Z_n} \varphi^c_A \widetilde{\Delta}^{AA} \varphi_{A}^{c} +  \varphi^c_B \widetilde{\Delta}^{BB} \varphi_{B}^{c}
	- \kappa \sum_{c,d \in \Z_n} \varphi^c_{\partial} T^{cd} \widetilde{\Delta}^{\partial\overline{\partial}} \varphi^d_{\overline{\partial}}
	\right).
	\label{eq:GS_Renyi_entropy_lattice_2}
\end{equation}
We can perform a discrete Fourier transform by the unitary change of variables $\psi^d = V_{c}^d \varphi^c$ withthe Vandemonde matrix
\begin{equation}
	V_d^c = \frac{1}{\sqrt{n}} \zeta_n^{cd}
	\label{eq:DFT_matrix}
\end{equation}
where $\zeta_n = e^{-2\pi i/n}$ is an $n$th root of unity.  Since $T$ is a circulant matrix, it is diagonalized by this coordinate transformation, giving
\begin{equation}
	V^\dagger T V = \operatorname{diag}\left( 1, \cos\left( \frac{2\pi}{n} \right), \cos\left(2 \frac{2\pi}{n} \right),\dots, \cos\left( (n-1) \frac{2\pi}{n} \right) \right).
	\label{eq:clockface_picture}
\end{equation}
The result of this transformation is that the $n$-sheeted partition function is now decoupled into $n$ single-sheet partition functions where, at the boundary between $A$ and $B$, the strength of the bonds between the two sides is between zero and one. Therefore, we find $\Tr \widetilde{\rho}_A^n = \prod_{k=0}^{n-1} \frac{\widetilde{Z}_k}{\widetilde{Z}_{M}}$ with
\begin{equation}
	\widetilde{Z}_k =  
	\int \mathcal{D} \varphi^c \exp\left( - \kappa \sum_{c=1}^n \varphi^c_A \widetilde{\Delta}^{AA} \varphi_{A}^{c} + \varphi^c_B \widetilde{\Delta}^{BB} \varphi_{B}^{c} + \cos\left( k \frac{2\pi}{n} \right) \varphi^c_\partial \widetilde{\Delta}^{\partial\overline{\partial}} \varphi_{\overline{\partial}}^{c} \right).
	\label{GS_Renyi_entropy_lattice_3}
\end{equation}
In particular, $\widetilde{Z}_0= \widetilde{Z}_{M}$ since the bond between the two sides is ``full strength'', and all the rest have weaker-than-normal bonds across the boundary.  (See Figure \ref{fig:lattice_geometry_decoupled}.)

We are thus left with the evaluation of the partition function of decoupled free bosons with ``weak links'' along the entanglement cut $\partial$. This is similar in spirit to the problem considered by Kane and Fisher who argued that a weak potential scatterer (``weak link'') in one dimensional systems of interacting fermions (Luttinger liquids) with repulsive interactions is a relevant perturbation in the renormalization group (RG) sense, so that it effectively cuts the 1D chain at low energy \cite{Kane1992a}. We emphasize however that a weak link for the fermions {\it does not} correspond to a weak link in the (free boson) Luttinger liquid theory, but rather takes the form of a boundary sine-Gordon perturbation after bosonization. Nevertheless, we expect a similar picture to hold in our case: the weak links with strength $0<\lambda = \cos \left(k \frac{2\pi}{n} \right)<1$ should flow under renormalization to conformally invariant boundary conditions, either $\lambda=0$ (decoupled systems) or $\lambda=1$ (`healed' uniform system).  To determine which way the RG flow goes, we can study the perturbative stability of these two fixed points. From the lattice analysis above, the perturbation to the homogeneous free boson action is 
\begin{equation}
	V = \cos\left( k \frac{2\pi}{n} \right) \varphi^c_\partial \widetilde{\Delta}^{\partial\overline{\partial}} \varphi_{\overline{\partial}}^{c} = \lambda \sum_{q} \varphi_{\ell,q}\varphi_{\ell+1,q},
\end{equation}
with $q$ the coordinate along the cut, which in the continuum limit becomes 
\begin{equation}
	V = \int_\partial dy \Big[ \lambda_1 \varphi (x=0,y)^2 + \lambda_2 \varphi (x=0,y) \partial_x \varphi (x=0,y) + \lambda_3 (\partial_x  \varphi (x=0,y))^2 + \cdots \Big]
\end{equation}
with $y$ the coordinate along the entanglement cut. (Recall we are working with a non-compact boson here.) Starting from an homogenous system with $\lambda=1$, we see that weakly weakening links along the cut is a strongly relevant perturbation that acts as a local mass term pinning down the field to $\varphi (x=0,y) = 0$ along the cut $\partial$ at low energy. This indicates that the homogeneous fixed point with $\lambda=1$ is unstable, and flows under renormalization to the free boson theory that vanishes (satisfying Dirichlet boundary conditions) along the cut (corresponding to $\lambda=0$). 

One can then self-consistently check that this fixed point with $\phi(x=0,y)=0$  is stable, as the perturbation $V$ around this fixed point $V_3 = \int_\partial dy \left( \lambda_3 (\partial_x \varphi (x=0,y))^2 + \dots \right)$ has scaling dimension $\Delta=2 > 1$ and is indeed irrelevant as boundary perturbation. For large systems, we may thus treat the $n-1$ sheets with weak links as if they had no link across the boundary, {\it i.e.} Dirichlet boundary conditions on $\partial$, consistent with the result \rev{(\textit{cf}. Equation \eqref{eq:ground_state_Renyi})} of Fradkin and Moore
\begin{equation}
	\Tr \widetilde{\rho}_A^n = \left( \frac{\widetilde{Z}_A \widetilde{Z}_B}{\widetilde{Z}_M} \right)^{n-1}.
	\label{eq:Renyi_entropies_lattice}
\end{equation}

Applying the replica trick, the entanglement entropy is
\begin{equation}
	\widetilde{S}_A = - \lim_{n\to 1} \frac{\partial}{\partial n} \Tr \widetilde{\rho}_A^n = \widetilde{F}_A + \widetilde{F}_B - \widetilde{F}_M.
	\label{eq:EE_GS_lattice}
\end{equation}
This is, of course, precisely the same as the continuum result \eqref{eq:S_QLM_GS_I} up to the topological term. One can again apply the result \eqref{eq:free_energy_scaling}, whose logarithmic term is universal, to conjecture that entropy takes the same form as in the continuum:
\begin{equation}
	\widetilde{S}_A = 2 f_s L - \frac{c}{6}\Big( \chi_A + \chi_B - \chi_{A\cup B} \Big) \ln L + {\cal O}(1).
	\label{eq:EE_GS_lattice_II}
\end{equation}
The appearence of the Euler characteristic comes from an application of the Gauss-Bonnet Theorem in the context of a manifold with smooth boundary. In the case of a shape with straight edges and sharp corners, like the rectangle of Figure \ref{fig:lattice_geometry}, there is an extra contribution that depends on the angles of the corners~\cite{Cardy:1988ic}
\begin{equation}
	\Delta \widetilde{F}(\gamma) = \frac{ \gamma}{24 \pi}\left( 1- \frac{\pi^2}{\gamma^2} \right) \ln L.
	\label{eq:angle_fcn_ground_state}
\end{equation}
Conical singularities on the boundary give a slightly different contribution. When writing the entanglement entropy, if the boundary $\partial$ between $A$ and $B$ has a corner of angle $\gamma$,  we must include the contribution from both the angle $\gamma$ and $2\pi-\gamma$. Corners on the boundary of $M$ are cancelled out and do not contribute. 

This can be explicitly confirmed in special cases with asymptotics on the lattice. David and Duplantier showed~\cite{Duplantier1988} that on a square lattice of size $(K-1) \times (L-1)$,
\begin{equation}
	\begin{aligned}
	&\ln \det \widetilde{\Delta}\\
	&= \frac{4 G}{\pi} K L - (K + L) \ln\left( \sqrt{2}+1 \right) - \frac{1}{4} \ln\left( KL \right) + \ln\left( \eta(e^{-2\pi \zeta}) \zeta^{1/4} \right) + \frac{5}{4} \ln 2 + \mathcal{O}\left( \frac{1}{KL} \right)
\end{aligned}
	\label{eq:D_D_lattice_asmptotic}
\end{equation}
\rev{where $G$ is Catalan's number, $\eta$  is the Dedekind $\eta$-function, and $\zeta = K/L$.} For the rectangular geometry of Figure \eqref{fig:lattice_geometry}, this implies that the logarithmic term in the entanglement has coefficient $-1/4$, which is consistent with \eqref{eq:angle_fcn_ground_state} \rev{for} 4 angles of $\pi/2$. The result \eqref{eq:D_D_lattice_asmptotic} was extended by Kenyon to any simply connected planar shape by mapping the problem to counting states of the classical dimer model \cite{Kenyon2000}. The connection is suggestive that perhaps this lattice regularization is closely linked to the quantum dimer model but beyond the scope of this paper.

\subsection{Numerical Confirmation}

\begin{figure}[h]
	\begin{tikzpicture}
		\node at (9,0){\includegraphics{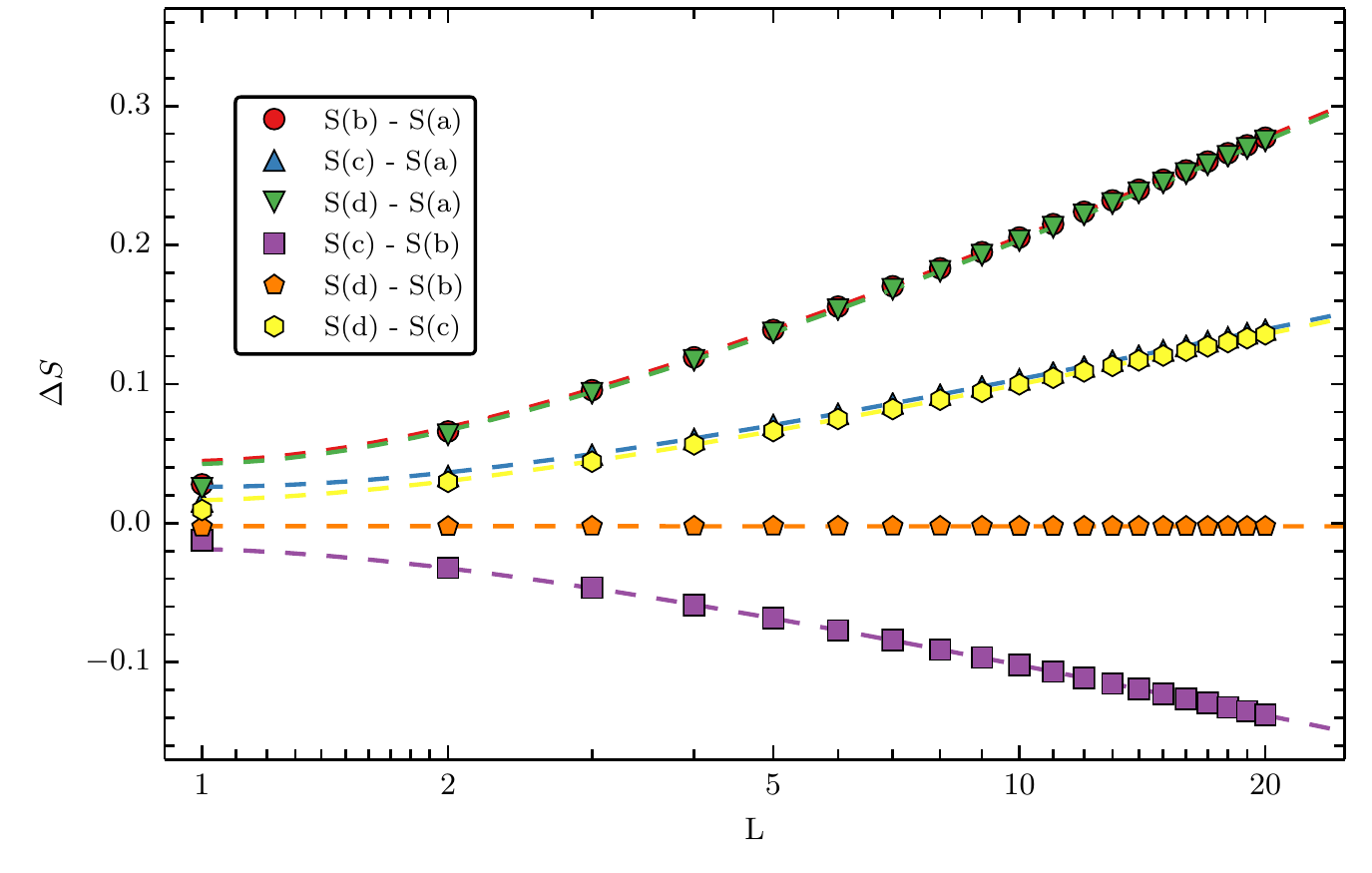}};
	\begin{scope}[scale=0.96,shift={(0,-3.4)}]
		\draw[thick] (0,0) rectangle (2,2);
		\draw[thick] (0,2) rectangle (2,4);
		\draw[thick] (0,4) rectangle (2,6);
		\draw[thick] (0,6) rectangle (2,8);
		
		\node at (0.4,7.6) {(a)};
		\node at (0.4,5.7) {(b)};
		\node at (0.4,3.7) {(c)};
		\node at (0.4,1.7) {(d)};

		\draw[thick,fill=black] (0.2,6.8) rectangle (1.8,7.2);
		\draw[thick,fill=black] (0.4,5.4) -- (1.6,5.4)
			-- (1.6,5.0)
			-- (1.2,5.0)
			-- (1.2,4.6)
			-- (0.8,4.6)
			-- (0.8,5.0)
			-- (0.4,5.0)
			-- cycle;
		\draw[thick,fill=black] (0.4,3.4) -- (1.6,3.4)
			-- (1.6,2.6)
			-- (1.2,2.6)
			-- (1.2,3.0)
			-- (0.4,3.0)
			-- cycle;
		\draw[thick,fill=black] (0.4,1.4) -- (1.2,1.4)
			-- (1.2,1)
			-- (1.6,1)
			-- (1.6,0.6)
			-- (0.8,0.6)
			-- (0.8,1)
			-- (0.4,1)
			-- cycle;
	\end{scope}
	
	\node at (4.3,4) {(e)};

	\end{tikzpicture}

	\caption{(a) -- (d) The four tetrominos. (e) Differences in entropy between tetrominos. Some differences have the same leading coefficient and are thus essentially superimposed. Fits for $L \ge 5$ (dashed lines) are described in the table below.}
	\label{fig:tetris_fits}
\end{figure}
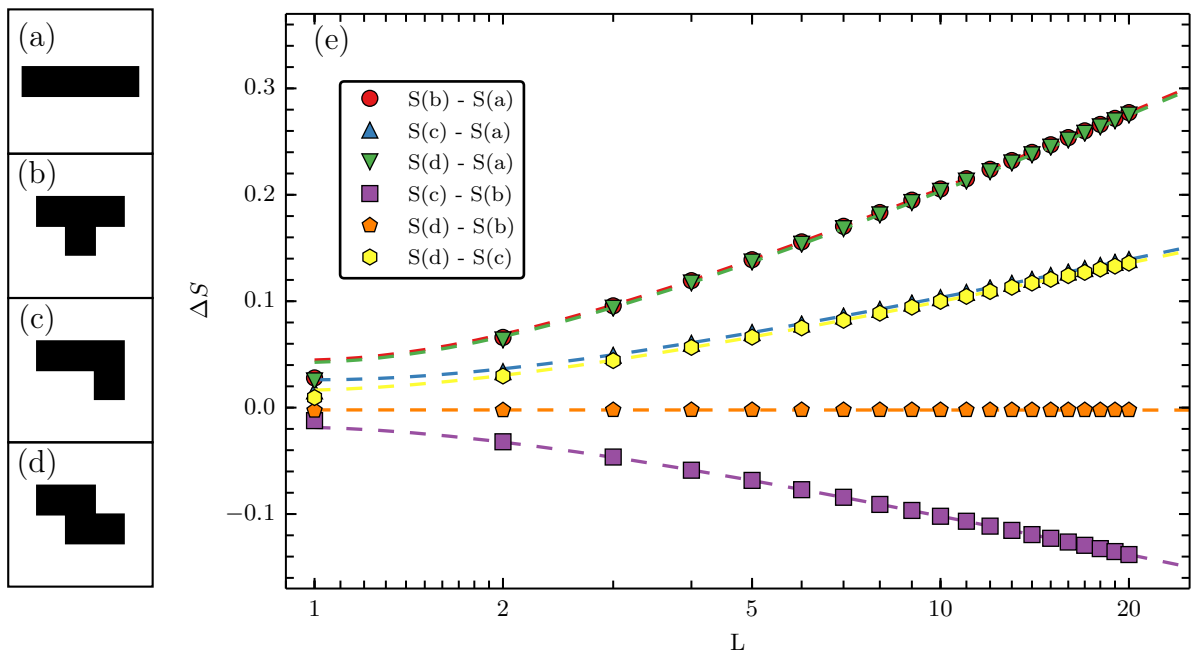

The discrete setting here gives us a chance to confirm \eqref{eq:angle_fcn_ground_state}. To numerically extract the sub-leading logarithmic behavior, it is necessary to peel back the linear term of \eqref{eq:EE_GS_lattice_II}. One way to do this is by considering the difference in entanglement between two regions $A_1, A_2 \subset M$ with the same area \textit{and} the same perimeter, but a different number of corners. A set of shapes with exactly this property are the \textit{tetrominos}, shown in Figure \ref{fig:tetris_fits}. (This is conceptually similar to methods to detect topological order in a ground state wavefunction~\cite{PhysRevLett.96.110404,PhysRevLett.96.110405}.) To do this, the entanglement entropy for the tetrominos shown in Figure \ref{fig:tetris_fits} was calculated by explicitly computing $\widetilde{F} = \frac{1}{2}\ln \det \widetilde{\Delta}$ on the lattice.

This was done for four tetrominos with area area of $4L^2$ and a perimeter of $10L$. Each entropy was evaluated via numerical determinants with the shape centered in a $12L \times 12L$ grid for $1 \le L \le 20$. Differences in the entropies of pairs are shown in Figure 4. These are fit to 
\begin{equation}
	S_i(L) - S_j(L) = c_1 \ln L + \frac{c_2}{L} + c_3,
	\label{eq:fit_equation}
\end{equation}
using linear least squares. There are finite size effects at small $L$, so fits only used $L \ge 5$. All fits had $R^2 > 0.99$. The results confirm \eqref{eq:EE_GS_lattice_II} to precision of $\sim10^{-4}$. From this demonstration and the above RG argument, we may conclude that the entanglement entropy in the ground state of the QLM model has the same form on the lattice as in the continuum. 

\begin{center}
\begin{tabular}[h]{c c *{4}{r}}
	\toprule
	\textbf{$S_i - S_j$} 
	& \textbf{Prediction $c_1$}
	& \textbf{Fit $c_1$}
	& \textbf{Fit $c_2$}
	& \textbf{Fit $c_3$}
	& \textbf{$c_1$ Error}\\
	\midrule
	$S_2 - S_1$ & $\tfrac{1}{9}$ 	& $0.111146$		& $0.105768$	& $-0.0610243$		& $0.35\times 10^{-4}$\\
	$S_3 - S_1$ & $\tfrac{1}{18}$	& $0.055652$		& $0.056490$	& $-0.0303294$		& $0.97\times 10^{-4}$\\
	$S_4 - S_1$ & $\tfrac{1}{9}$	& $0.111045$		& $0.105610$ 	& $-0.0630068$		& $-0.66\times 10^{-4}$\\
	$S_3 - S_2$ & $-\tfrac{1}{18}$	& $-0.055497$		& $-0.049279$ 	& $0.0306949$		& $0.62\times 10^{-4}$\\
	$S_4 - S_2$ & $0$				& $-0.000101$		& $-0.000158$ 	& $-0.0019825$		& $-1.01\times 10^{-4}$\\
	$S_4 - S_3$ & $\tfrac{1}{18}$	& $0.055393$		& $0.049120$ 	& $-0.0326775$		& $-1.63\times 10^{-4}$\\
	\bottomrule	
\end{tabular}
\end{center}

\section{Lattice Excited State Entanglement}
\label{sec:lattice_excited}

\rev{The previous section showed that the entanglement entropy in the ground state is the same on the lattice and in the continuum, up to non-universal contributions. It is reasonable to expect that this exact correspondence carries over to excited states as well. Indeed this is the case. This section parallels Section \ref{sec:continuum_excited_states} to find arbitrary R\'enyi entropies, but on the lattice. The correspondence is so close that many of the details are repetitive and are given in \ref{app:Renyi_lattice_excited}. Here, however, all definitions can be made precise and all quantities are divergence-free and finite. This allows explicit numerical computations of the EPAs. We conclude with a physical interpretation of the EPAs.}


Consider the excited lattice wavefunctional
\begin{equation}
	\widetilde{\Psi_\lambda}[\{\varphi_{pq}\}] = \widetilde{Z}_{M}^{1/2} \sqrt{2\kappa \widetilde{\lambda}} \varphi^{\lambda}_{pq} e^{-\frac{1}{2} \widetilde{S}[\varphi_{pq}]} = \widetilde{Z}_{M}^{1/2} \sqrt{2\kappa \widetilde{\lambda}} \sum_{(pq) \in {M}} \widetilde{L}^\lambda_{pq} \varphi_{pq} e^{-\frac{1}{2} \widetilde{S}[\varphi_{pq}]},
	\label{eq:lattice_excited_wavefunctional}
\end{equation}
the direct analogue of \eqref{eq:excited_wavefunctional} on the lattice. Here $\widetilde{L}^\lambda$ is the eigenvector of the lattice Laplacian on ${M}$ with eigenvalue $\widetilde{\lambda}$.

Unlike the ground state, where the $A$ and $B$ sides contributed independently to the entanglement, here the two sides are explicitly linked. This can be seen already by examining the reduced density matrix. By definition, for the state $\widetilde{\Psi_\lambda}$ it is 
\begin{equation}
	\begin{aligned}
		&\braket{\varphi_A^1|\widetilde{\rho}_A\left(\widetilde{\Psi_\lambda}  \right)|\varphi_A^2}\\
		\ &=\ \int \mathcal{D} \varphi_B^1 \; \left( \sum_M \widetilde{L^\lambda}_M \varphi^{11}_M \right) \left( \sum_M \widetilde{L^\lambda}_M \varphi^{21}_M \right)	\exp\left( - \frac{\kappa}{2} \sum_{M} \varphi_{M}^{11} \widetilde{\Delta}^{M} \varphi_{M}^{11} +  \varphi_{M}^{21} \widetilde{\Delta}^{M} \varphi_{M}^{21} \right),
	\end{aligned}
	\label{eq:reduced_density_excited_lattice}
\end{equation}
where $\varphi^{21}$ takes the values of $\varphi^2$ on $A$ and $\varphi^1$ on $B$, and similarly for $\varphi^{11}$. As in the previous section, the path integral is a notational shortcut for the integrals of $\varphi_{pq}$ over all lattice sites and \textit{not} a literal path integral. After computing the Gaussian integrals one finds
\begin{align}
	\label{eq:reduced_density_excited_lattice_computed_a}
	&\braket{\varphi_A^1|\rho_A|\varphi_A^2}\\
	\label{eq:reduced_density_excited_lattice_computed_b}
	\ &=\ \Big[ \varphi_A^1 \varphi_A^2 + \frac{1}{2}\left( \varphi_A^1 + \varphi_A^2 +2\right) (\varphi_A^1 + \varphi_A^2) \widetilde{\Delta}^{\partial \overline{\partial}} \widetilde{G}^{BB} \widetilde{L^\lambda}_B + \widetilde{L^\lambda}_B \widetilde{G}^{BB}  \widetilde{L^\lambda}_B \Big]\\
	\label{eq:reduced_density_excited_lattice_computed_c}
	 &\qquad \times\ Z_B \exp\left( 
	 \frac{1}{2} \varphi_A^1 \widetilde{\Delta}^{AA} \varphi_A^1 
	 +\frac{1}{2} \varphi_A^2 \widetilde{\Delta}^{AA} \varphi_A^2
	 +\frac{1}{8}\left( \varphi_A^1 + \varphi_A^2 \right) \widetilde{\Delta}^{\partial \overline{\partial}} \widetilde{G}^{BB} \widetilde{\Delta}^{\overline{\partial}\partial} \left( \varphi_A^1 + \varphi_A^2 \right)
	 \right)
\end{align}
where $\widetilde{G}^{BB}$ is the inverse of the discrete Laplacian on $B$, and as above $ \widetilde{\Delta}^{\partial \overline{\partial}}$ is the restriction of the discrete Laplacian to the boundary, which acts as a ``tunneling'' matrix between the right-most column of $A$ and the left-most column of $B$. The first and last terms of \eqref{eq:reduced_density_excited_lattice_computed_b} depend purely on the $A$ side or the $B$ side and will become the analogues of $\alpha$ and $\beta$ in the R\'enyi entropy. The second term involves tunneling from one side to the other and will manifest as the \textit{inter}-sheet term $\gamma$. It is therefore not sufficient to consider separately the density matrix restricted to the $A$ and $B$ sides, but rather one must consider the behavior on and across the boundary as well. Indeed, the boundary behavior is crucial in what follows. This is in contrast to the behavior in the ground state, where the two sides contributed to the entanglement independently. 

The R\'enyi entropies for excited states on the lattice involve
\begin{equation}
	\Tr \widetilde{\rho}_A^n = \frac{1}{\widetilde{Z}_M} \int \mathcal{D}[\varphi^c] \prod_{c \in \Z_n} \left( \sum_M \widetilde{L^\lambda}_M \varphi^{cc}_M \right) \left( \sum_M \widetilde{L^\lambda}_M \varphi^{(c+1)c}_M \right) e^{-\widetilde{S}[\varphi_M^c]}.
	\label{eq:lattice_Renyi_excited}
\end{equation}
It is again sufficient to compute 2-point functions on the $n$-sheeted geometry, then apply Wick's theorem and integrate against the eigenvectors of the Laplacian. This is carried out explicitly in \ref{app:Renyi_lattice_excited}. The upshot is one can write all the R\'enyi entropies of $\widetilde{\Psi}_\lambda$ in terms of lattice EPAs
\begin{subequations}
\begin{align}
	\label{eq:alpha_lattice}
	\widetilde{\alpha} \ &=\ \widetilde{\lambda} a^2 \sum_{(pq),(rs) \in A} \widetilde{L}^\lambda_{pq}  \widetilde{G}^A_{pqrs} \widetilde{L}^\lambda_{rs},\\
	\label{eq:beta_lattice}
	\widetilde{\beta} \ &=\ \widetilde{\lambda} a^2\sum_{(pq),(rs) \in B} \widetilde{L}^\lambda_{pq}  \widetilde{G}^B_{pqrs} \widetilde{L}^\lambda_{rs},\\
	\label{eq:gamma_lattice}
	\widetilde{\gamma} \ &=\ \widetilde{\lambda} a^2\sum_{(pq),(rs) \in M} \widetilde{L}^\lambda_{pq}  \left[\widetilde{G}^M_{pqrs} - \widetilde{G}^D_{pqrs} \right] \widetilde{L}^\lambda_{rs}.
\end{align}
\end{subequations}
These are the direct analogues of \eqref{eq:a} -- \eqref{eq:d}. Since $\widetilde{L}^\lambda$ has dimensions of inverse length, the factors of $a^2$ is necessary to preserve dimensions. 

\rev{These may be simplified.} By definition, 
$\widetilde{\Delta} \widetilde{L}^\lambda = - \widetilde{\lambda} \widetilde{L}^\lambda$ and $\widetilde{\Delta}^M \widetilde{G}^M = \text{Id}^M$. What one would like to do is use the factors of $\widetilde{L}^\lambda$ to produce extra Laplacians, to get
\begin{equation}
	\widetilde{\gamma} = \frac{a^2}{\widetilde{\lambda}} \sum_M \widetilde{L}^\lambda\left( \widetilde{\Delta}^M  \widetilde{G}^M  \widetilde{\Delta}^M - \widetilde{\Delta}^M \widetilde{G}^D \widetilde{\Delta}^M \right) \widetilde{L}^\lambda,
	\label{eq:gamma_lattice_II}
\end{equation}
then reduce the Green's functions to Kronecker deltas. Unfortunately, this is not possible because the action of the Laplacian does not commute with restricting the domain, so $\widetilde{\Delta}^M \widetilde{G}^A \neq \text{Id}^A$ (resp. $B$ and $D$). Nevertheless, one can do this away from the boundary. As previously, let $\DDD$ be the Laplace operator restricted to the \textit{links} between $A$ and $B$, i.e. the tunneling operator between $\partial$, the sites on $A$ closest to $B$ and $\overline{\partial}$, the sites on $B$ closest to $A$ (see Figure \ref{fig:lattice_geometry}). Then
\begin{align}
	\widetilde{\Delta}^M \widetilde{G}^D \ &=\ \left( \widetilde{\Delta}^D + \DDD \right) \widetilde{G}^D = \text{Id}^M + \DDD \widetilde{G}^D,\\
	\widetilde{\Delta}^M \widetilde{G}^D \widetilde{\Delta}^M 
	\ &=\ \widetilde{\Delta}^M + \DDD + \DDD \widetilde{G}^D \DDD.
\end{align}
Combining these and \eqref{eq:gamma_lattice_II},
\begin{equation}
	\widetilde{\gamma} = \frac{a^2}{\widetilde{\lambda}} \sum_{\partial, \overline{\partial}} 
	\begin{bmatrix}
		\widetilde{L}^\lambda_\partial & \widetilde{L}^\lambda_{\overline{\partial}}
	\end{bmatrix}
	\begin{bmatrix}
		\DDD \widetilde{G}^B_{\overline{\partial} \overline{\partial}} \DDD & \DDD\\[0.5em]
		\DDD & 	\DDD \widetilde{G}^A_{\partial \partial} \DDD\\
	\end{bmatrix}
	\begin{bmatrix}
		\widetilde{L}^\lambda_\partial\\[0.5em]
		\widetilde{L}^\lambda_{\overline{\partial}}
	\end{bmatrix}.
	\label{eq:explicit_gamma_lattice}
\end{equation}

As a concrete example, consider the $K \times L$ rectangular geometry where $A$ is an $\ell \times L$ rectangle, as shown in Figure \ref{fig:lattice_geometry}. Numerically, $\DDD$ is minus the identity matrix on $L$ elements, which gives
\begin{equation}
	\widetilde{\gamma} = 	
\frac{a^2}{\widetilde{\lambda}} \sum_{q,s=1}^{L}
	\begin{bmatrix}
		\widetilde{L}^\lambda_{\ell,q} & \widetilde{L}^\lambda_{\ell+1,q}
	\end{bmatrix}
	\begin{bmatrix}
		\widetilde{G}^B_{\ell+1,q,\ell+1,s} & - \delta_{qs}\\
		-\delta_{qs} & \widetilde{G}^A_{\ell,q,\ell,s}\\
	\end{bmatrix}
	\begin{bmatrix}
		\widetilde{L}^\lambda_{\ell,s}\\
		\widetilde{L}^\lambda_{\ell+1,s}
	\end{bmatrix}
	\label{eq:explicit_gamma_lattice_rectangle}.
\end{equation}

\begin{figure}

	\begin{tikzpicture}[scale=2]
		\node at (4.75,0.3) {\includegraphics{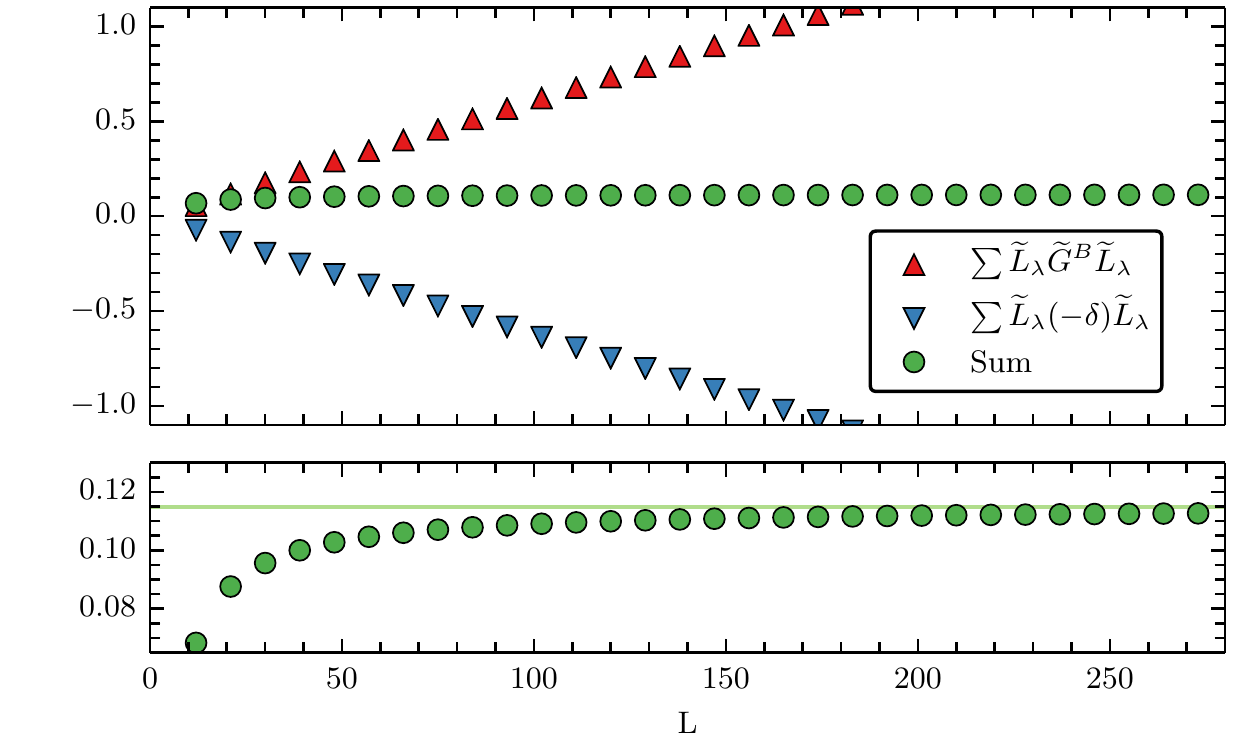}};
		\begin{scope}[shift={(0,1.15)}]
		\draw (0,0) rectangle (1.9,1);
		\draw[dashed] (0.9,0) -- (0.9,1);
		\draw[thick, middlearrow={latex}] (1.0,0.6) -- (0.8,0.6);
		\node at (0.3,0.75) {(a)};
		\node at (2.8,0.75) {(d)};
		\end{scope}

		\begin{scope}[shift={(0,0.025)}]
		\draw (0,0) rectangle (1.9,1);
		\draw[dashed] (0.9,0) -- (0.9,1);
		\draw[postaction={on each segment={middlearrow={latex}}},-,thick] (0.8,0.8) 
				-- (1.2,0.8) 
				-- (1.2,0.6) 
				-- (1.6,0.6)
				-- (1.6,0.8)
				-- (1.8,0.8)
				-- (1.8,0.2)
				-- (1.2,0.2)
				-- (1.2,0.4)
				-- (1.0,0.4)
				-- (1.0,0.2)
				-- (0.8,0.2);
		\node at (0.3,0.75) {(b)};
		\end{scope}
i
		\begin{scope}[shift={(0,-1.1)}]
		\draw (0,0) rectangle (1.9,1);
		\draw[dashed] (0.9,0) -- (0.9,1);
		\draw[postaction={on each segment={middlearrow={latex}}},-,thick] (0.8,0.6) 
				-- (1.0,0.6)
				-- (1.0,0.8)
				-- (1.8,0.8)
				-- (1.8,0.2)
				-- (1.0,0.2)
				-- (1.0,0.6);
		\node at (0.3,0.75) {(c)};
		\node[fill=white] at (2.8,0.75) {(e)};
		\end{scope}
	\end{tikzpicture}
	\caption{Lattice scattering processes. (a)~``Tunneling'' across the boundary. (b)~Tunneling across the boundary, scattering on the $B$ side, and tunneling back. (c)~A tadpole-like contribution contributing to the divergences. (d)~The contributions due to processes (a) and (b) and their sum. Data is for the mode $k_x = 4, k_y= 3$ as described in \ref{app:boundary_explicit_geometry}.3. (e)~Their sum is $\mathcal{O}(1)$ and quickly converges as $L \to \infty$.}
		\label{fig:lattice_scattering}
\end{figure}
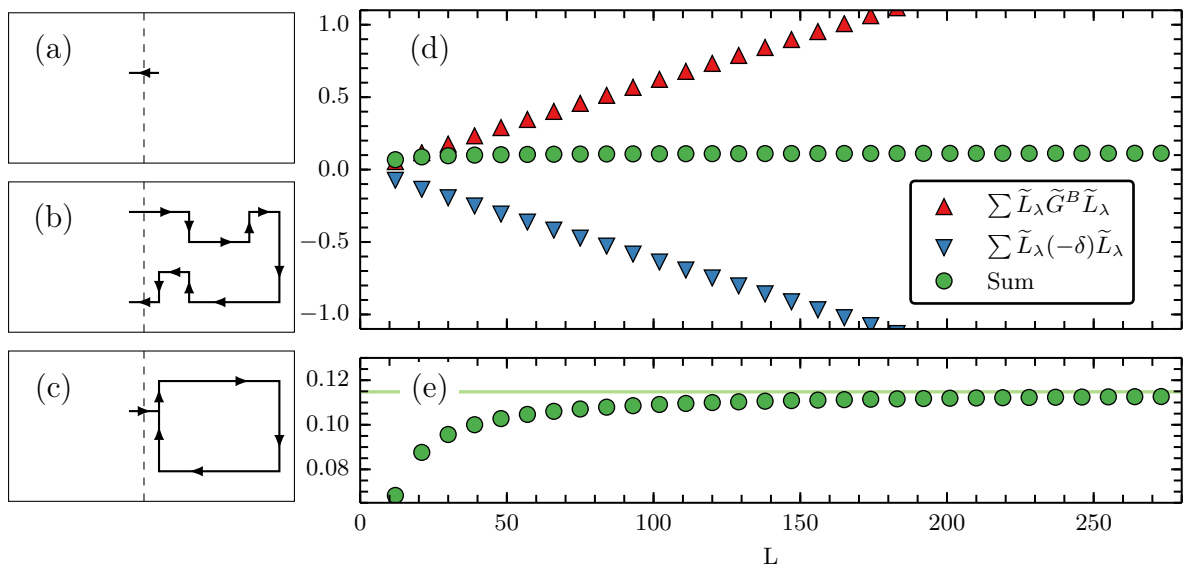

\rev{
	The lattice form for the EPAs suggests a physical interpretation in terms of scattering processes. We stress that this is an analogy suggested by the form of the equations to promote intuition and \textit{not} a genuine physical scattering process. For $\widetilde{\gamma}$ we may consider each entry of the matrix in \eqref{eq:explicit_gamma_lattice_rectangle} as a scattering process.  The upper right entry, $-\delta_{qs}$ represents ``tunneling'' from the $\ell$th row to the $\ell+1$th row (Figure \ref{fig:lattice_scattering} (a)). Meanwhile, the upper left entry, $\widetilde{G}^B_{\ell+1,q,\ell+1,s}$, represents starting on the $\ell$th row, tunneling across, scattering around anywhere in $B$, and tunneling back across the boundary (Figure \ref{fig:lattice_scattering} (b)). Hence, $\widetilde{\gamma}$ is the sum over all scattering from the $A$ to $B$, minus the scattering back --- a ``current'' flowing across the boundary. Alterntively, $\widetilde{\gamma}$ can be interpreted as the scattering on $M$ sourced by $\widetilde{L}^\lambda$ minus the scatter strictly within $A$ and within $B$, which leaves the scattering across $\partial$. Though similar logic, one may interpret $\widetilde{\alpha}$ as the scattering within the $A$ side, minus what scatters out through $\partial$, and likewise for $B$.

	It is instructive to examine the scaling of these separate processes. The ``tunneling'' process scales as the number of lattice sites along the boundary, i.e. it diverges as ${\cal O}(L)$. However, for each lattice site on the edge, there is a ``tadpole'' process that tunnels across at a site and then tunnels back at the same site, shown in (Figure \ref{fig:lattice_scattering} (c)). The weight of this also diverges (due to the logarthmic divergence of the Green's function) as ${\cal O}(L)$ with equal but opposite coefficient to the tunneling process. In the thermodynamic limit, therefore, the divergences precisely cancel and $\widetilde{\gamma}$ is a constant. Visually, processes (a) and (c) in Figure \ref{fig:lattice_scattering} cancel each other out and leave only the (b) processes which ``tunnel across and then tunnel back''.  This is shown numerically for the rectangular geometry in Figure \ref{fig:lattice_scattering} (d). Thus we may conclude that the inter-sheet EPA, $\widetilde{\gamma}$, is generically non-zero and comparable in magnitude to $\widetilde{\alpha}$ and $\widetilde{\beta}$. By applying conformal mappings, this holds whenever $A$ and $B$ are simply connected and it is reasonable that the same conclusions extend to more general domains.
}

\section{Continuum Redux}
\label{sec:continuum_redux}

As with the ground state, the lattice and continuum results are closely analogous. The same structure of cancelling divergences also appears in the continuum and thus necessitates careful regularization of any integrals involved, which may be quite subtle. One can think of the lattice results as a guide for interpreting the continuum quantities.

A key tool to do this is Green's second identity, which we recall here for convenience: suppose $f: U \subset \R^2 \to \R$ and $G^U$ is the Green's function on $U$ with Dirichlet boundary conditions (so $\Delta^U G^U = \text{Id}^U$), then for $w \in U$,
\begin{align}
	\int_U d^2z \; \Delta_z^U f(z) G^U(z,w) &= \int_U d^2z \; f(z) \Delta^U_x G^D(z,w) \notag\\
	&+ \int_{\partial U} dS_z\; G^U(z,w) \frac{\partial}{\partial n_z} f(z) - f(z) \frac{\partial}{\partial n_z} G^U(z,w),
	\label{eq:green_second_identity}
\end{align}
where $\frac{\partial}{\partial n_z} = \hat{n} \cdot \nabla_z$ is the outwards-pointing normal derivative with respect to $z$.

For our case, $G^U(z,w) = 0$ identically when $w \in \partial U$, so this reduces to
\begin{equation}
	\int_U d^2z \; \Delta_z^U f(z) G^U(z,w) =
	f^U(w)
- \int_{\partial U} dS_z \; f(w) \frac{\partial}{\partial n_z} G^U(z,w),
	\label{eq:green_second_identity_II}
\end{equation}
where $ f^U(w) = f(w)$ if $w \in U \setminus \partial U$ and $0$ if $w \in \partial U$. The crucial point here is that if $f(w)$ does not vanish on the boundary, then the values of $f$ on the boundary are not preserved by $\text{Id}^U$, which only acts as the identity for functions which vanish on $\partial U$; the information on the boundary is lost. 

Let us apply this to $\alpha$. Since $\Delta^M L_\lambda + \lambda L_\lambda=0$,
\begin{equation}
	\alpha = \lambda \int_M d^2z \, \int_M d^2w \, L_\lambda(z) G^A(z,w) L_\lambda(y) = \frac{1}{\lambda} \int_A d^2z\, \Delta_z L_\lambda(z) \int_A d^2w\, \Delta_w L_\lambda(w) G^A(w,z).
\end{equation}
Applying \eqref{eq:green_second_identity_II} for both $z$ and $w$ yields, after a calculation, 
\begin{equation}
	\begin{aligned}
	&\alpha = \int_A d^2z \, L_\lambda(z)\\
	&\quad \underbrace{- \frac{1}{\lambda} \int_\partial dS_z\, L_\lambda(z) \frac{\partial}{\partial n_z} L_\lambda^A(z) + \frac{1}{\lambda} \int_\partial dS_z\, \int_\partial dS_w \, L_\lambda(z) \frac{\partial}{\partial n_z} \frac{\partial}{\partial n_w} G^A(z,w) L_\lambda(w)}_{=-d_A}.
\end{aligned}
\end{equation}
Let $d_B$ be the same as $d_A$ with $A \leftrightarrow B$, so
\begin{equation}
	\alpha = \int_A d^2z \, L_\lambda^2(z) - d_A, \quad \beta = \int_B d^2z \, L_\lambda^2(z) - d_B,\quad \gamma = d_A + d_B.
	\label{eq:alpha_beta_gamma_continuum}
\end{equation}
Then $\alpha + \beta + \gamma = \int_A d^2z \, L_\lambda^2(z) + \int_B d^2z \, L_\lambda^2(z) = 1$ due to the fact that $L_\lambda$ is normalized. 

The interpretation of $\int_A d^2z \, L_\lambda^2(z)$ and (and $A\leftrightarrow B$) is clear, but $d_A$ and $d_B$ are strange expressions: they involve the normal derivatives of $L_\lambda^A$ at the boundary of $A$ where $L_\lambda^A$ is discontinuous. The double normal derivative of the Green's function on the boundary is subtle because the Green's function vanishes at the boundary except when $z=w$ where it is singular. This hypersingular integral kernel arises in several other places, such as in the engineering literature of fracture analysis and always requires careful regularization \cite{Sutradhar2008}.

From the lattice, we are motivated to interpret these quantities as
\begin{align}
	\frac{\partial}{\partial n_z} L_\lambda^A
	\ &=\ \lim_{\varepsilon \to 0} \frac{L_\lambda^A(z) - L_\lambda^A(z-\varepsilon \hat{n})}{\varepsilon}
	=\lim_{\varepsilon \to 0} -\frac{L_\lambda^A(z-\varepsilon)}{\varepsilon}\\
	\frac{\partial}{\partial n_z} \frac{\partial}{\partial n_w} G^A(z,w) 
	\ &=\ \lim_{\varepsilon \to 0} \frac{G^A(z-\varepsilon, w-\varepsilon)}{\varepsilon^2}.
	\label{eq:interpretations_of_continuum_quantities}
\end{align}
Here $\hat{n}$ is the outwards pointing unit normal, $z,w \in \partial$, and $z-\varepsilon$ is a point $\varepsilon$ inside the boundary. So our regulated version is
\begin{equation}
	d_A = \frac{1}{\lambda} \lim_{\varepsilon\to 0} \left(\int_\partial dS_z L_\lambda(z) \left[ \frac{L_\lambda(z-\varepsilon)}{\varepsilon} \right] + \int_\partial dS_z \, \int_\partial dS_w \, L_\lambda(z) \left[ \frac{G^A(z-\varepsilon,w-\varepsilon)}{\varepsilon^2} \right] L_\lambda(w)\right).
	\label{eq:d_A_regulated}
\end{equation}
Both terms of this are, of course, superficially divergent quantities. But just as on the lattice, the divergences cancel to leave a finite quantity. Roughly this is because the normal derivative of the Green's function restricted to the boundary is a delta function, so 
\begin{equation}
	\frac{G(z-\varepsilon,w-\varepsilon)}{\varepsilon^2} \sim - \frac{\delta(z-w)}{\varepsilon} + \mathcal{O}(1).
	\label{eq:green_function_near_bondary}
\end{equation} 
A more precise version of this is shown in \ref{app:boundary_explicit_geometry}. Because of the minus sign, the $\mathcal{O}(\varepsilon^{-1})$ terms cancel
\begin{equation}
	\begin{aligned}
	d_A &\sim \frac{1}{\lambda} \lim_{\varepsilon\to 0}\left( \int_\partial dS_z L_\lambda(z) \left[ \frac{L_\lambda(z-\varepsilon)}{\varepsilon} \right]  \right.\\
	&\qquad \left. +  \int_\partial dS_z \, \int_\partial dS_w \, L_\lambda(z) \left[ - \frac{\delta(z-w)}{\varepsilon} + \mathcal{O}(1) \right] L_\lambda(w)\right) = \mathcal{O}(1).
\end{aligned}
	\label{eq:cancelling_divergences}
\end{equation}

This regulation scheme means $\alpha$, $\beta$, and $\gamma$ are well-defined and finite for any choice of geometry. In principle they can be calculated whenever the Green's function is known. In practice, this requires either an explicit equation for the Green's function on both $A$ and $B$. When the boundary is smooth, there is a general form for the $\mathcal{O}(1)$ in \eqref{eq:green_function_near_bondary} in terms of Hadamard regularized integrals. Just as in Cardy and Peschel's expression for the free energy~\cite{Cardy:1988ic}, there appear to be special cases for corners, which generalize \eqref{eq:angle_fcn_ground_state} in some sense.

\begin{figure}[h]
		\begin{center}
			\includegraphics{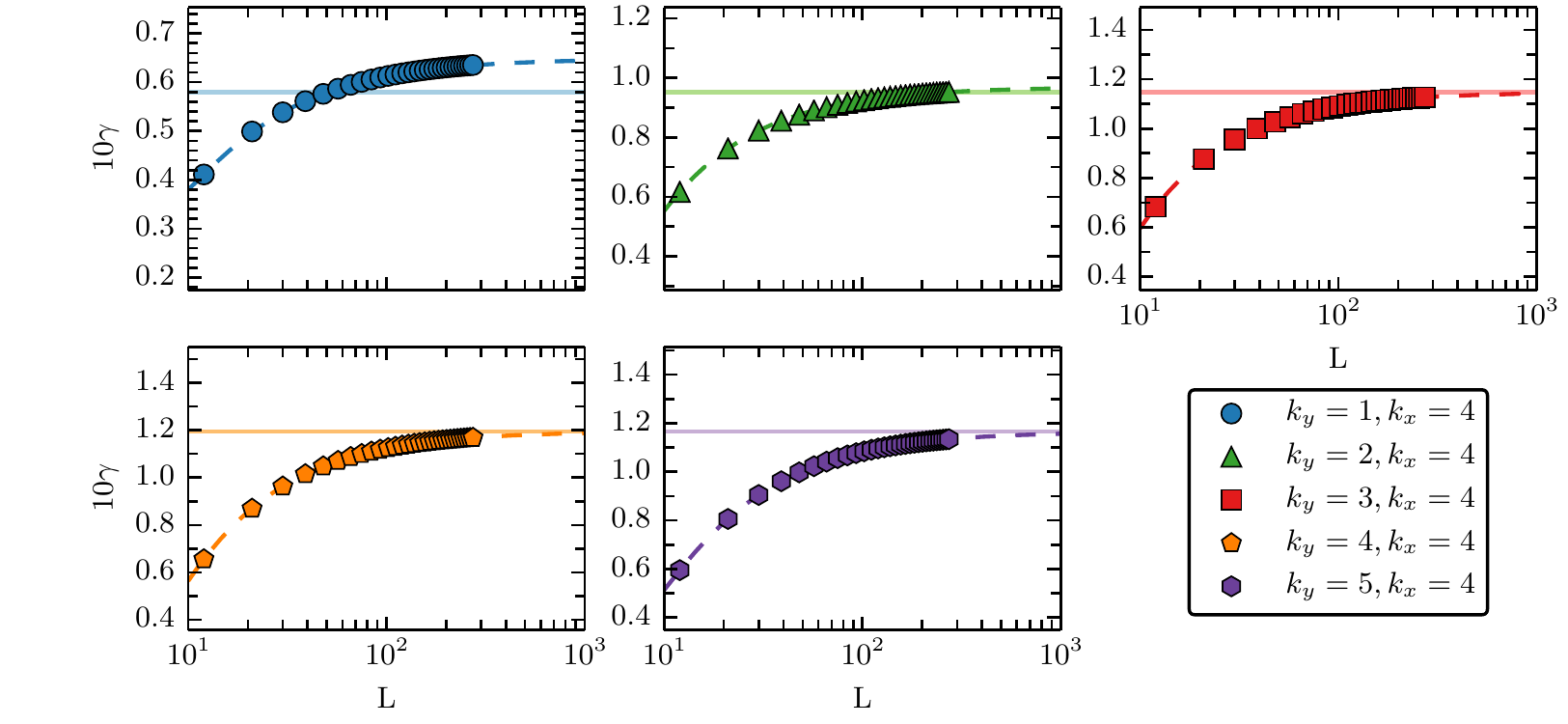}\\
		\end{center}
		\caption{Comparison of the EPAs on the lattice (points and dashed fits) and in the continuum (solid lines) in the rectangular geometry. Details of the geometry and fits are given in \ref{app:boundary_explicit_geometry}.3. The agreement is only good for large $k_y$ due to approximations made in the continuum calculation of $\gamma$.}
		\label{fig:d_universal_rect}
	\end{figure}

\begin{figure}[h]
		\begin{center}
			\includegraphics{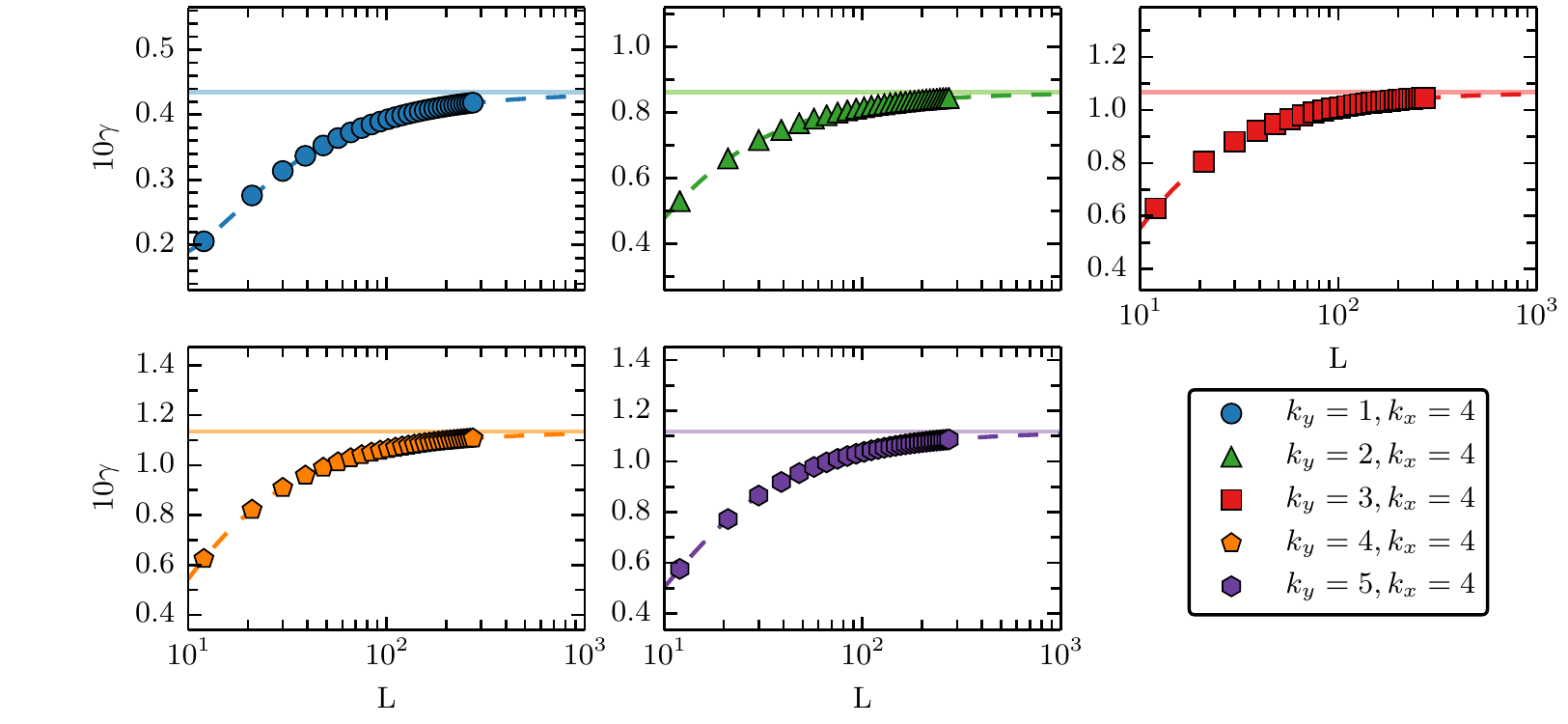}
		\end{center}
		\caption{Comparison of the EPAs on the lattice (points and dashed fits) and in the continuum (solid lines) in the artificial half-plane geometry. Details of the geometry and fits are given in \ref{app:boundary_explicit_geometry}.3.}
		\label{fig:d_universal_hp}
	\end{figure}

As an example, consider the rectangle of side lengths $L_x$ and $L_y$ with an entanglement cut at $x = \ell_x$. The modes of the Laplacian are precisely Fourier modes in this geometry,
\begin{equation}
	L_{k_x,k_y}(x,y) = \frac{2}{L_x L_y} \sin \frac{\pi k_x x}{L_x} \sin \frac{\pi k_y y}{L_y},
	\label{eq:rect_modes}
\end{equation}
with $\lambda = \left( \frac{\pi k_x}{L_x} \right)^2 + \left( \frac{\pi k_y}{L_y} \right)^2$. In \ref{app:boundary_explicit_geometry}, it is shown that 
\begin{align}
	\alpha \ &=\ \frac{\ell_x}{L_x} - \frac{1}{2 \pi k_x} \sin \left(2 \pi k_x\frac{\ell_x }{L_x} \right) - d^\text{rect},\\
	\beta \ &=\  \frac{L_x - \ell_x}{L_x} + \frac{1}{2 \pi k_x} \sin \left(2 \pi k_x\frac{\ell_x }{L_x} \right) - d^\text{rect},\\
	\gamma \ &=\ 2d^\text{rect},
\end{align}
where, for $L_x = L_y = L$,
\begin{equation}
	d^\text{rect} \approx \frac{8}{\pi^3} \frac{2 \pi k_y \mathcal{S}(2 \pi k_y) - \pi k_y \mathcal{S}(\pi k_y) - (-1)^{ky} + 1 }{k_x^2 + k_y^2} \sin \left(  \pi k_x \frac{\ell_x}{L} \right)^2.
	\label{eq:d_rectangle}
\end{equation}
Here $\mathcal{S}(x) = \int_0^x \frac{\sin t}{t} \, dt$ is the sin integral. The approximations made, described in \ref{app:boundary_explicit_geometry}, are accurate for large $k_y$. Figure \ref{fig:d_universal_rect} shows the comparison between numerical calculations of $\gamma$ on the lattice and \eqref{eq:d_rectangle}, where good agreement can be seen for large $k_y$. One can also consider an artificial half-plane geometry with no corners, in which case the agreement is essentially perfect, shown in Figure \ref{fig:d_universal_hp}. This is described in detail in \ref{app:boundary_explicit_geometry}. This is strong evidence that EPAs are numerically equal in the continuum and on the lattice, and are hence universal.

\section{Conclusions}

In this paper we studied the entanglement of excited states in the quantum Lifshitz model. We found that there is an infinite class of excited states $\Psi_\lambda$ (reaching arbitrarily high energies) whose entanglement is completely specified by three geometrical, albeit state-dependent, ``entanglement propagator amplitudes'' $\alpha$, $\beta$, and $\gamma$ as defined in Equation \eqref{eq:a} -- \eqref{eq:d}. In terms of these, the second R\'enyi entropy is particularly simple:
\begin{equation}
	S_2\left( \Psi_\lambda \right) = - \ln \Tr \rho_A^2 = S_2(\text{GS}) - \ln \left(a^2 + b^2 + g^2 \right),
	\label{eq:second_Renyi_one_quantum}
\end{equation}
where $a = \alpha + \gamma/2$, $b = \beta + \gamma/2$ and $g = \gamma/2$. A combinatorial formula for arbitrary R\'enyi entropies is given in Eqution \eqref{eq:Renyi_entropies_alpha_beta_gamma}. This same procedure works for any state, but the combinatorics is generally complex. A particularly simple case are the states $\Psi_{K_\lambda}$ with $K$ quanta in mode $\lambda$, whose second R\'enyi entropy has the multinomial-esque expression
\begin{equation}
	S_2(\Psi_{K_\lambda}) =  S_2(\text{GS}) - \ln \left( \sum_{k_1,k_2 = 1}^K  \left[ \binom{K}{k_1 k_2} \, a^{k_1} b^{k_2} g^{K - k_1 - k_2} \right]^2 \right).
	\label{eq:second_Renyi_n_quanta}
\end{equation}

The EPAs $\alpha$, $\beta$, and $\gamma$ have a natural interpretation as probabilities since they are all between zero and one and sum to unity. It is quite possible that these constants, which can be defined purely from Green's functions of the Laplacian, may have already been studied in the mathematics literature under another name. The lattice results of Section \ref{sec:lattice_excited} suggest that one can think of $\gamma$ as the total ``flow'' across the boundary and as $\alpha$ as the ``energy'' on the $A$ side, minus what flows out, and similarly for $\beta$. The continuum results of Section \ref{sec:continuum_redux} show that these quantities are geometric but contain subtle cancelling divergences. Naively, since $\gamma$ can be defined as a quantity within a lattice site of the boundary, it appears to be a microscopic quantity. However, the comparison between lattice and continuum results gives strong evidence that $\gamma$ gives a leading order universal contribution to the entanglement. 

As a by-product of working on the lattice, Section \ref{sec:lattice_GS} reconstructed the ground state entanglement entropy on the lattice \eqref{eq:EE_GS_lattice}. Previous derivations of this fact \cite{Zaletel:2011dz,Zhou2016a} made use of careful changes of variables in path integrals, whereas the lattice derivation relies renormalization group arguments.  One can therefore consider this as independent confirmation of the non-topological part of the result \eqref{eq:S_QLM_GS_I}.

In the special case where $\alpha = \beta$, we may use the replica trick \cite{Calabrese:2004hl} to determine the entanglement entropy. Starting from \eqref{eq:all_n_Renyi}, and differentiating yields
\begin{equation}
	S(\Psi_\lambda; \alpha = \beta) = S(\text{GS}) +\ln2 - \sum_{k=2}^\infty \frac{(-1)^k \Big( \Gamma(k+1,-1)+\Gamma(k,-1) \Big)}{e\, \Gamma(k) \,\Big(\Gamma(k)+\Gamma(k-1)\Big)} \; \gamma^k,
	\label{eq:entanglement_entropy}
\end{equation}
where $\Gamma$ is the incomplete gamma function. The infinite sum converges quickly, even when $\gamma=1$, and is $\mathcal{O}(1)$. With a sufficient application of combinatorics, it is undoubtedly possible to extend this to $\alpha \neq \beta$ and perhaps even to more general states. In all excited states with one quantum, the difference with the ground state entanglement is fully determined by the EPAs and is therefore $\mathcal{O}(1)$ and universal. 

The notion of EPAs is entirely general and, in principle, works for any state whose wavefunctional is known. For instance, an eigenstate which populates $M$ modes can be written down in terms of $3\binom{M+1}{2}$ generalizations of $\alpha$, $\beta$ and $\gamma$ such as \eqref{eq:alpha_lambda_mu}. Of course, more complicated wavefunctions have more complex combinatorics, making it harder to extract information. \rev{One} possible direction would be to find a state with a particularly simple wavefunctional and examine the entanglement dynamics after a global or local quench. This was studied in the special case of a local quench by a vertex operator in \cite{Zhou2016}. \rev{The ``propagating'' nature of the EPAs is reminiscent of the light-cone picture of entanglement spreading, but since our results are at finite time, we see no connection between the two at present.}

Another future direction is to examine entanglement at finite temperature. Since entanglement entropy is not a good measure of entanglement in mixed states, it is standard in field theory to instead calculate the logarithmic negativity which is an entanglement monotone and bounds several other common measures of entanglement \cite{Vidal2002}. It may be computed with a modified replica trick~\cite{Calabrese:2014bs} which amounts to sightly modifying the combinatorics. At low temperature where only a few modes are populated, it may be possible to find an analytic continuation and determine the negativity. At high temperatures, one may be able to leverage the strong mathematical results on the asymptotic distribution of eigenvalues of the Laplacian into an asymptotic expression for the entanglement.

\paragraph{Acknowledgments.}

We thank J. Cardy for helpful discussions and Adam Scherlis for suggesting tetrominos. We acknowledge support from the Department of Energy through the LDRD program of LBNL (RV), from NSF DMR-1507141 and a Simons Investigatorship (JEM), and center support from CaIQuE and the Moore Foundation's EPiQS initiative.

\appendix

\section{A Modified Wick's Theorem for the Many-Sheeted Geometry}
\label{app:wick_thm}

This Appendix will derive Equation \eqref{eq:wick_modified}, a version of Wick's Theorem suitably modified for fields defined on the many-sheeted surface $\mathcal{S}_n$. The first uses a series of coordinate transformations introduced in \cite{Zaletel:2011dz} to decouple the sheets and the second evaluates the path integrals. The derivation has been written to be clear at the expense of concision.

Recall we are dealing with a compactified field $\varphi$, so $\varphi \equiv \varphi + 2 \pi R$ for some $R >0$. We can therefore rewrite the ``surface expectation'' \eqref{eq:surface_expectation} concretely as a constrained path integral
\begin{equation}
	\braket{\mathcal{O}(x_1) \mathcal{O}(x_2)\cdots }_S = Z_M^{-n} \int_{\text{con}} \mathcal{D} \varphi_a \left( \mathcal{O}(x_1) \mathcal{O}(x_2)\cdots \right) e^{-\sum_{a=1}^n S[\varphi_a]}.
	\label{eq:surface_expectation_1}
\end{equation}
where $\int_\text{con}$ means that the fields obey the constraint 
\begin{equation}
	\varphi_a(x) = \varphi_b(x) + 2\pi R \omega_{ab}, \quad \forall x \in \partial, \omega_{ab} \in \Z.
	\label{eq:con1}
\end{equation}
First, let us make the change of variables $\varphi_a = \varphi_{\alpha_a} + \widetilde{\varphi}_a$ where the first term $\varphi_{\alpha_a}$ is topological and encodes the zero modes and vortices, obeying $\Delta \varphi_{\alpha_a} = 0$ with Dirichlet boundary conditions on $M$. The new field $\widetilde{\varphi}_a$ is non-compact. By applying Green's first identity, the action decouples: $S[\varphi_a] = S[\varphi_{\alpha_a}] + S\left[ \widetilde{\varphi}_a \right]$, and
\begin{equation}
	\int_\text{con} \mathcal{D}[\varphi_a] = \sum_{\alpha_a} \int_{\text{con}'} \mathcal{D}[\widetilde{\varphi}_a]
	\label{eq:con_path_integral}
\end{equation}
where the sum is over all possible zero modes and $\text{con'}$ is the constraint
\begin{equation}
	\widetilde{\varphi}_a(x) - \widetilde{\varphi}_b(x)  = \varphi_{\alpha_a} - \varphi_{\alpha_b} +  2\pi R \omega_{ab}, \quad \forall x \in \partial, \omega_{ab} \in \Z, \forall \alpha_a.
	\label{eq:con2}
\end{equation}
Adopting a matrix notation, one may express this as $M \v{\widetilde{\varphi}} = \mathbf{c}_w(x)$ where one may choose (other choices are possible)
\begin{equation}
	M = \begin{pmatrix}
		-1 & 1 & 0 & \cdots & 0\\
		0 & -1 & 1 & \cdots & 0\\
		\vdots & \vdots & \vdots & \ddots & \vdots\\
		1 & 0 & 0 & \cdots & -1\\
	\end{pmatrix}
	\text{ and } \mathbf{c}(x) = \begin{pmatrix}
		\varphi_{\alpha_1} - \varphi_{\alpha_2} +  2\pi R \omega_1\\
\varphi_{\alpha_2} - \varphi_{\alpha_3} +  2\pi R \omega_2\\
\vdots\\
\varphi_{\alpha_n} - \varphi_{\alpha_1} +  2\pi R \omega_n\\
	\end{pmatrix}
	\label{eq:con_matrix_form}
\end{equation}
where $w_k \in \Z$ and $\sum_k w_k = 0$. Note that $\Delta_{a_k} \varphi(x_k) = \Delta_{a_k} \left( \varphi_{\alpha_a} + \widetilde{\varphi}_a \right) = \Delta_{a_k} \tilde{\varphi}_a$, so a product of $\varphi$'s is equal to the same product of $\widetilde{\varphi}$'s, plus terms that are harmonic in at least one of the variables. These additional terms will not concern us because they can be removed with an integration by parts, and will be denoted $\text{Har}$. So after this first change of coordinates, the surface expectation is 
\begin{equation}
	\begin{aligned}
		&\braket{\varphi_{a_1}(x_1)\cdots \varphi_{a_{2N}}(x_{2N})}_{\mathcal{S}_n}\\
	\ &=\ Z_M^{-n} \sum_{\alpha_a,w} e^{-\sum_a S[\varphi_{\alpha_a}]} \int_{\setc{M \v{\widetilde{\varphi}}=\mathbf{c}_w(x)}{x \in \partial}}  \mathcal{D}[\tilde{\varphi}_a] \; 
	\widetilde{\varphi}_{a_1}(x_1)\cdots \widetilde{\varphi}_{a_{2N}}(x_{2N}) e^{-\sum_a S[\tilde{\varphi}_a]} + \text{Har}.
\end{aligned}
	\label{eq:surface_expectation_2}
\end{equation}

We now perform a second change of coordinates to decouple the sheets. Choose some $T \in O(n)$ such that $T_1^a = n^{-1/2}, \forall a \in \Z_n$, and put $\v{\psi} = T \widetilde{\varphi}$. There are many possible such matrices $T$. One explicit family \cite{Zaletel:2011dz} is
\begin{equation}
	T = \begin{pmatrix}
		\frac{1}{\sqrt{2}} & - \frac{1}{\sqrt{2}} \\
		\frac{1}{\sqrt{6}} & \frac{1}{\sqrt{6}} & \frac{-2}{\sqrt{6}}\\
		\frac{1}{\sqrt{12}} & \frac{1}{\sqrt{12}} & \frac{1}{\sqrt{12}} & \frac{-3}{\sqrt{12}}\\
		\vdots & \vdots & \vdots & \vdots & \ddots\\
		\frac{1}{\sqrt{n(n-1)}} & 
		\frac{1}{\sqrt{n(n-1)}} & 
		\frac{1}{\sqrt{n(n-1)}} & 
		\frac{1}{\sqrt{n(n-1)}} & 
		\dots & 
		\frac{-(n-1)}{\sqrt{n(n-1)}}\\
		\frac{1}{\sqrt{n}} &
		\frac{1}{\sqrt n} &
		\frac{1}{\sqrt n} &
		\frac{1}{\sqrt n} &
		\dots &
		\frac{1}{\sqrt n}\\
	\end{pmatrix}
	\label{eq:T_matrix}.
\end{equation}
Since the action is quadratic in $\widetilde{\varphi}$, $S[\widetilde{\varphi}_a] = S[\psi_a]$, and the constraint becomes
\begin{equation}
	TMT^{-1} \v{\psi} = T \mathbf{c}(x).
	\label{eq:con3}
\end{equation}
This has the special property that $\psi_1$ is unconstrained, and the constraint on the other $(n-1)$ fields is an invertible matrix. In particular, they are constrained by
\begin{equation}
	\psi_{a>1} = \left( TM^{-1} \mathbf{c} (x) \right) = c'(x)_{a>1}, \quad x \in \partial.
	\label{eq:con4}
\end{equation}
Therefore
\begin{equation}
	\begin{aligned}
		&\braket{\varphi_{a_1}(x_1)\cdots \varphi_{a_{2N}}(x_{2N})}_\text{con}\\
	&= Z_M^{-n} \sqrt{n}^{-L/a} \sum_{\alpha_a,w} e^{-\sum_a S[\varphi_{\alpha_a}]} 
	\int \mathcal{D}[\tilde{\varphi}'_1]
	\int_{\setc{\psi_{a>1}\partial =  c_{a>1}'(x)}{x \in \partial}} \mathcal{D}[\psi_{a>1}]
	\; e^{-\sum_a S[\psi_a]}\\
	&\qquad \times \sum_{b_1, b_2, \cdots, b_{2N} = 1}^n \left( T^{-1} \right)_{a_1}^{b_1} \psi_{b_1}(x_1)\cdots \left( T^{-1} \right)_{a_{2N}}^{b_{2N}} \psi_{b_{2N}}(x_{2N}) + \text{Har}
\end{aligned}
	\label{eq:surface_expectation_3}
\end{equation}
The factor of $\sqrt{n}^{-L/a}$ comes from the Jacobian determinant and is non-universal.

To make it possible to actually evaluate the path integrals, we make one last coordinate transformation. For $a =2$ to $n$, let $C_a$ be the solution to the Dirichlet problem $C_a = c(x)$ on $\partial$, $C_a = 0$ on $\partial M$, and $\Delta C_a = 0$ on $M \setminus \partial$. This specifies $C_a(x)$ uniquely on $M$. Then define $\psi'_a = \psi_a + C_a$ for $a>1$ and $\psi'_1 = \psi_1$. Again note that $\Delta \psi'_a = \Delta \psi_a$, so the integrands of the path integrals are the same up to harmonic terms, and the actions again decouple, so
\begin{equation}
	\begin{aligned}
		&\braket{\varphi_{a_1}(x_1) \cdots \varphi_{a_2N}(x_{2N})}_{\mathcal{S}_n}\\
		\ &=\ Z_M^{-n} \left( n^{-L/2a} \sum_{\alpha_a, w} e^{-\sum_a S[\varphi_{\alpha_a}] - \sum_{a>1} S[C_a]} \right) \int_M \mathcal{D}[\psi'_1] \int_D \mathcal{D}[\psi'_a] \\
		\ &\qquad \times \sum_{b_1,b_2,\dots,b_{2N} =1}^n  \left( T^{-1} \right)_{a_1}^{b_1} \psi'_{b_1}(x_1) \cdots \left( T^{-1} \right)_{a_{2N}}^{b_{2N}} \psi'_{b_{2N}}(x_{2N}) e^{-\sum_a S[\psi'_a]} + \text{Har},
	\end{aligned}
	\label{eq:surface_expectation_4}
\end{equation}
where $\int_M$ means there is no constraint on $\partial$ and $\int_D$ means that the fields obey the \textit{Dirichlet} boundary condition $\psi_a' = 0$ on $\partial$. At this point, the topological part, in parentheses, is completely separate from the path integrals. This is the end of the first part: the sheets have been decoupled and the path integrals may now be evaluated.

The path integrals now describe expectations of a free boson with two sets of boundary conditions: either nothing happens at $\partial$, or the fields are set to zero there. As in the above paragraph, these will be referred to as $M$ or $D$ respectively. Let us recall the standard (normalized) Wick's theorems for these geometries
\begin{equation}
	\begin{aligned}
		&\braket{\psi'(x_1)\cdots \psi'(x_{2k})}_{M/D}\\
		&\hspace{2em} = Z_{M/D}^{-1} \int_{M/D} \mathcal{D} [\psi']\;  \left( \psi'(x_1)\cdots \psi'(x_{2k}) \right) e^{-S[\psi']} = \left( 2\kappa \right)^{-k} \sum_{\text{``pairs"}} G^{M/D}(x_{c_i},x_{d_i})
	\end{aligned}
	\label{eq:Wick_thm_standard}
\end{equation}
where $Z_D = Z_A Z_B$, $G^M$ is the Green's function for the Laplacian on $M$ with Dirichlet boundary conditions and if $G^A$ and $G^B$ are the corresponding Green's functions for $A$ and $B$, then
\begin{equation}
	G^D(x,y) = \begin{cases}
		G^A(x,y) & \text{if $x, y \in A$}\\
		G^B(x,y) & \text{if $x, y \in B$}\\
		0 & \text{else}.
	\end{cases}
\end{equation}
In the $D$ case, the fields are constrained to zero along $\partial$, so correlations may not be ``transmitted'' through $\partial$, but they can for the $M$ case. The sum over ``pairs'' may be written explicitly as the sum over the set of all possible partitions of $2k$ integers:
\begin{equation}
	\text{``pairs''} = \setc{C = (c_1,\dots, c_k),D=(d_1,\dots,d_k) \subset \Z_{2k}}{C\cup D = \Z_{2k}}.
	\label{eq:pairs_as_set_app}
\end{equation}
This is necessary to keep track of the explicit indices below.

The first field $\psi'_1$ is evaluated with the ``$M$'' Wick's theorem, and the rest use the ``$D$'' version. Since the fields are now decoupled, the expectation of the product of $\psi'_a$'s is the product of the expectations. Hence
\begin{equation}
	\begin{aligned}
		&\braket{\varphi_{a_1}(x_1) \cdots \varphi_{a_{2N}}(x_{2N})}_{\mathcal{S}_n} =  \left( n^{-L/2a} \sum_{\alpha_a, w} e^{-\sum_a S[\varphi_{\alpha_a}] - \sum_{a>1} S[C_a]} \right)\\ 
	\ &\qquad\times\	\sum_{b_1,b_2,\dots,b_{2N}=1}^n \left( 2\kappa \right)^{-N}\left( \prod_k \left( T^{-1} \right)_{a_k}^{b_k} \right) \frac{Z_M}{Z_M} \Braket{\prod_{S^1 = \setc{k}{b_k=1}} \psi'_1(x_k)}_M\\
		\ &\qquad \times\ \frac{Z_D}{Z_M} \Braket{\prod_{S^2 = \setc{k}{b_k=2}} \psi'_2(x_k)}_D \times  \cdots \times \frac{Z_D}{Z_M} \Braket{\prod_{S^2 = \setc{k}{b_k=n}} \psi'_n(x_k)}_D + \text{Har},
	\end{aligned}
	\label{eq:surface_expectation_5}
\end{equation}
This serves as a definition for $S^a$. The overall prefactor is the ground state R\'enyi entropy
\begin{equation}
	\Tr \rho_A^n(\text{GS}) =   \left( n^{-L/2a} \sum_{\alpha_a, w} e^{-\sum_a S[\varphi_{\alpha_a}] - \sum_{a>1} S[C_a]} \right)\\
 \left( \frac{Z_D}{Z_M} \right)^{n-1},
\end{equation}
where the first parenthetical is topological. All that remains is to rearrange terms to simplify the combinatorics:
\begin{align*}
	& \ \frac{\braket{\varphi_{a_1}(x_1) \cdots \varphi_{a_{2N}}(x_{2N})}_{\mathcal{S}_n}}{\left( 2\kappa \right)^{-N} \Tr \rho_A^n(\text{GS})} - \text{Har}\\
	\ &=\ \sum_{b_1,\dots,b_{2N}=1}^n \left( \prod_{k=1}^{2N} \left( T^{-1} \right)_{a_k}^{b_k} \right) \Braket{\prod_{S^1 = \setc{k}{b_k=1}} \psi'_1(x_k)}_M \prod_{a=2}^n \Braket{\prod_{S^a = \setc{k}{b_k=a}} \psi'_a(x_k)}_D\\
	\ &=\ \sum_{b_1,\dots,b_{2N}=1}^n \left( \prod_{k=1}^{2N} \left( T^{-1} \right)_{a_k}^{b_k} \right) \left( \sum_{C^1 \cup D^1 = S^1} \left( 2\kappa \right)^{-\n{S^1}} G^D(x_{c_1^1}, x_{d_1^1}) G^D(x_{c_2^1},x_{d_2^1}) \cdots  \right)\\
	\ &\qquad \times\ \prod_{a=2}^n  \left( \sum_{C^a \cup D^a = S^a} \left( 2\kappa \right)^{-\n{S^a}} G^D(x_{c_1^a}, x_{d_1^a}) G^D(x_{c_2^a},x_{d_2^a}) \cdots  \right)\\
	\intertext{where the sums $C^a$ and $D^a$ run over all possible pairing of elements of $S^a$. The sums of all possible pairings for each field is the same as the sum over all possible pairings of all the fields, restricted so that the two fields in each pair are the same}
	\ &=\  \sum_{b_1,\dots,b_{2N}=1}^n \left( \prod_{k=1}^{2N} \left( T^{-1} \right)_{a_k}^{b_k} \right) \sum_{C\cup D = \Z_{2N}} \prod_{i=1}^N \delta_{b_{c_i}b_{d_i}} \begin{cases}
		G^M(x_{c_i},x_{d_i}) & \text{if $b_{c_i} = b_{d_i} = 1$}\\
		G^D(x_{c_i},x_{d_i}) & \text{if $b_{c_i} = b_{d_i} \neq 1$}\\
	\end{cases}\\
	\intertext{We can now eliminate the $T$'s. Since these sums are finite, everything commutes, so}
	\ &=\ \sum_{C\cup D = \Z_{2N}} \prod_{i=1}^{N} \sum_{b_{c_i},b_{d_i}=1}^n \delta_{b_{c_i}b_{d_i}} 
	\left( T^{-1} \right)^{b_{c_i}}_{a_{c_i}} \left( T^{-1} \right)^{b_{d_i}}_{a_{d_i}}
	\begin{cases}
		G^M(x_{c_i},x_{d_i}) & \text{if $b_{c_i} = b_{d_i} = 1$}\\
		G^D(x_{c_i},x_{d_i}) & \text{if $b_{c_i} = b_{d_i} \neq 1$}\\
	\end{cases}\\
	\intertext{Eliminating the Kronecker delta}
	\ &=\ \sum_{C\cup D = \Z_{2N}} \prod_{i=1}^{N} \left(\sum_{b=1}^n \left( T^{-1} \right)_{a_{c_i}}^b \left( T^{-1} \right)_{a_{d_i}}^b G^D(x_{c_i},x_{d_i})\right. \\
	&\qquad\ \left.+ \left( T^{-1} \right)_{a_{c_i}}^1  \left( T^{-1} \right)_{a_{d_i}}^1 \left[G^M(x_{c_i},x_{d_i}) - G^D(x_{c_i},x_{d_i}) \right] \right)
	\intertext{By definition, $T^{a}_1 = n^{-1/2}$ and $T$ \rev{is} orthogonal, so $T^{-1} = T^t$. Therefore}
	\ &=\ \sum_{C\cup D = \Z_{2N}} \prod_{i=1}^{N} \left(\sum_{b=1}^n \left( T \right)^{a_{c_i}}_b \left( T^{-1} \right)_{a_{d_i}}^b G^D(x_{c_i},x_{d_i}) + \frac{1}{n} \left[G^M(x_{c_i},x_{d_i}) - G^D(x_{c_i},x_{d_i}) \right] \right)\\
	\ &=\ \sum_{C\cup D = \Z_{2N}} \prod_{i=1}^{N} \left( \delta_{a_{c_i}a_{d_i}} G^D(x_{c_i},x_{d_i}) + \frac{1}{n} \left[G^M(x_{c_i},x_{d_i}) - G^D(x_{c_i},x_{d_i}) \right] \right).
\end{align*}
This implies Equation \eqref{eq:wick_modified} for $N=n$.

\section{Arbitrary R\'enyi Entropies on the Lattice}
\label{app:Renyi_lattice_excited}

In this Appendix we evaluate excited state entropies on the lattice. By definition, the $n$th R\'enyi entropy is $\widetilde{S}_A^{(n)} = \frac{1}{1-n} \ln \Tr \widetilde{\rho}_A^n$. We must compute
\begin{equation}
	\Tr \widetilde{\rho}_A^n = \frac{1}{\widetilde{Z}_M} \int \mathcal{D}[\varphi^c] \prod_{c} \left( \sum_M \widetilde{L^\lambda} \varphi^{cc}_M \right) \left( \sum_M \widetilde{L^\lambda} \varphi^{(c+1)c}_M \right) e^{-\widetilde{S}[\varphi_M^c]}.
\end{equation}
 The calculation proceeds in close analogy to the continuum case in Section \ref{sec:continuum_excited_states}. It is again sufficient to compute 2-point functions on the $n$-sheeted geometry, then apply Wick's theorem and integrate against the eigenvectors of the Laplacian.

 First, apply the the discrete Fourier Transform $\psi^{c} = V_d^c \varphi^d$ from \eqref{eq:DFT_matrix}, the $n$-sheets are decoupled as described in the previous section. The RG arguments from Section \ref{sec:lattice_GS} may now be applied, whose result is that the fields $\psi_2, \psi_3,\dots, \psi_n$ have Dirichlet boundary conditions at $\partial$, so two-point functions on those sheets can only scatter from $A$ to $A$ or from $B$ to $B$, each according to the Dirichlet Green's function for $A$ or $B$ respectively. Meanwhile, $\psi_1$, the center-of-mass sheet, has nothing to distinguish $\partial$, so scattering proceeds by the Green's function on all of $M$. Therefore
\begin{equation}
	\begin{aligned}
	\frac{\braket{\psi^c(x) \psi^d(y)}_{\mathcal{S}_n}}{(2\kappa)^n \Tr \widetilde{\rho}_A^n(GS)} &= \delta^{cd} \left[ \delta_c^1 \widetilde{G}^M(x,y) + \left( 1-\delta_c^1) \widetilde{G}^D(x,y) \right) \right]\\
	&= \delta^{cd}\left[ \widetilde{G}^D(x,y) + \delta_c^1 \left( \widetilde{G}^M(x,y) - \widetilde{G}^D(x,y) \right) \right].
\end{aligned}
	\label{eq:two_point_fourier_space}
\end{equation}
Here $\widetilde{G}^M$ is the Dirichlet Green's function for the lattice Laplacian on $M$ and 
\begin{equation}
	\widetilde{G}^D(pq,rs) = \begin{cases}
		\widetilde{G}^A(pq,rs) & (pq),(rs) \in A\\
		\widetilde{G}^B(pq,rs) & (pq),(rs) \in B\\
		0 & \text{else.}
	\end{cases}
	\label{eq:dirichlet_lattice_green_fcn}
\end{equation}
We can thus compute the two-point function back on the original sheets
\begin{align*}
	\frac{\braket{\varphi^a(x) \varphi^b(y)}}{(2\kappa)^n \Tr \widetilde{\rho}_A^n(GS)}
	= V^{a}_c V^b_d \braket{\psi^c \psi^d}
	= \left( V V^\dagger \right)_a^b \widetilde{G}^D + V^a_1 V^b_1 \left( \widetilde{G}^M-\widetilde{G}^D \right).
\end{align*}
From the definition of $V$ and the fact it is unitary, it follows that
\begin{equation}
	\frac{\braket{\varphi^a(x) \varphi^b(y)}}{(2\kappa)^n \Tr \widetilde{\rho}_A^n(GS)}
	= \delta^{ab} \widetilde{G}^D(x,y) + \frac{1}{n}\Big[ \widetilde{G}^M(x,y) - \widetilde{G}^D(x,y) \Big].
	\label{eq:lattice_two_point_fcn}
\end{equation}
Physically, the inter-sheet term term occurs because the ``center of mass'' Fourier mode connects the $A$ side to the $B$ side. If you start with a field on sheet $c$, then $1/\sqrt{n}$ is part of the center of mass sheet, which may scatter to the other side, and then $1/\sqrt{n}$ of that is on sheet $d$. Unlike in the ground state, we cannot cancel out the inter-sheet term with the normalization, because here it is additive instead of multiplicative. 

Because this is a Gaussian theory (even on the $n$-sheeted surface), one may apply Wick's theorem and find that $2N$-point functions are
\begin{equation}
	\begin{aligned}
	&\frac{\braket{\varphi^{a_1}(x_1) \varphi^{a_2}(x_2) \cdots \varphi^{a_{2N}}(x_{2N})}_{S_N}}{(2\kappa)^n \Tr \widetilde{\rho}_A^n(GS)}\\
	&\hspace{2em}=  
		\sum_{\text{``pairs"}} 
		\prod_{i=1}^N
		\left( \delta_{a_{c_i} a_{d_i}}  \widetilde{G}^D(x_{c_i}, x_{d_i}) + \frac{1}{n} \left[ \widetilde{G}^M(x_{c_i},x_{d_i}) - \widetilde{G}^{D}(x_{c_i},x_{d_i}) \right] \right).
	\end{aligned}
		\label{eq:wick_modified_lattice}
\end{equation}
One can check that this is precisely the form of Equation \eqref{eq:wick_modified} under the replacement $\widetilde{G} \leftrightarrow G$ (without the topological prefactor for the ground state entropy, since we have not compactified on the lattice). By the same reasoning as Section \ref{sec:continuum_excited_states},
\begin{equation}
	\Tr \widetilde{\rho}_A^n\left(\Psi_\lambda \right)\\
	\ =\ \Tr \widetilde{\rho}_A^n(\text{GS}) \times \sum_{\text{``pairs"}} \prod_{i=1}^n  \frac{\widetilde{\gamma}}{n} +
\left.	\begin{cases}
	\widetilde{\alpha} & \text{if $c_i + 1 \equiv  d_i  \pmod n$ and $c_i$ even}\\
	\widetilde{\beta} & \text{if $c_i + 1 \equiv  d_i  \pmod n$ and $c_i$ odd}\\
		0 & \text{otherwise}
	\end{cases}
\right\}.
\label{eq:Renyi_entropies_alpha_beta_gamma_lattice}
\end{equation}
where
\begin{align}
	\widetilde{\alpha} \ &=\ \tilde{\lambda} a^2 \sum_{(pq),(rs) \in A} \widetilde{L}^\lambda_{pq}  \widetilde{G}^A_{pqrs} \widetilde{L}^\lambda_{rs},\\
	\widetilde{\beta} \ &=\ \tilde{\lambda} a^2\sum_{(pq),(rs) \in B} \widetilde{L}^\lambda_{pq}  \widetilde{G}^B_{pqrs} \widetilde{L}^\lambda_{rs},\\
	\widetilde{\gamma} \ &=\ \tilde{\lambda} a^2\sum_{(pq),(rs) \in M} \widetilde{L}^\lambda_{pq}  \left[\widetilde{G}^M_{pqrs} - \widetilde{G}^D_{pqrs} \right] \widetilde{L}^\lambda_{rs}.
\end{align}

\section{Boundary Term in Explicit Geometry}
\label{app:boundary_explicit_geometry}

\subsection{Behavior of the Green's function near the boundary}

A key quantity to evaluate the probabilities $\alpha,\beta,\gamma$ is the Green's function of the Laplacian with Dirichlet boundary conditions very close to the boundary (entanglement cut). Consider the Green's function on the half-plane,
\begin{equation}
	G\left( (x,y), (x_0,y_0) \right) = \frac{1}{4\pi}\ln \frac{(x-x_0)^2 + (y-y_0)^2}{(x-x_0)^2 + (y+y_0)^2}.
\end{equation}
Denote the Green's function a distance $\varepsilon = x = x_0 \ll 1$ away from the boundary by $G_\epsilon(u)$, with $u = \n{y - y_0}$ the coordinate along the boundary,
\begin{equation}
	G_\epsilon(u) = \frac{1}{4 \pi} \ln \frac{u^2}{u^2 + 4 \epsilon^2}.
\label{eqGnearBdry}
\end{equation}
It is easy to show that the normal derivative at the boundary is simply a delta function 
\begin{equation}
\partial_n G = \lim_{\epsilon \to 0} \frac{G_\epsilon(u)}{\epsilon} = - \delta(u),
\end{equation}
but we are actually interested in the subleading contribution to this delta function
\begin{equation}
\frac{G_\epsilon(u)}{\epsilon^2} = - \frac{\delta(u)}{\epsilon} + g(u) + {\mathcal O} (\epsilon), 
\label{eqExpG}
\end{equation}
where the divergent contribution $- \frac{\delta(u)}{\epsilon}$ will cancel out another delta function contribution in \eqref{eq:d_A_regulated}. We would like to understand what $g(u)$ is (as a distribution). Since the logarithm is hard to deal with, it is convenient to consider the Taylor expansion of the derivative instead $\frac{1}{\epsilon}\frac{d G_\epsilon(u)}{d \epsilon} = - \frac{\delta(u)}{\epsilon} +2 g(u) + {\mathcal O} (\epsilon)$, which differs from~\eqref{eqExpG} by a factor of 2.  
Let us compute this quantity explicitly. It reads:
\begin{equation}
\frac{1}{\epsilon}\frac{d G_\epsilon(u)}{d \epsilon} = -\frac{2}{\pi(u^2 + 4 \epsilon^2)},
\end{equation}
that we conveniently rewrite as
\begin{equation}
-\frac{2}{\pi(u^2 + 4 \epsilon^2)} = - \frac{16 \epsilon^2}{\pi (u^2 + 4 \epsilon^2)^2} + \frac{2}{\pi} \frac{d}{du} \left( \frac{u}{u^2 + 4 \epsilon^2}\right). 
\end{equation}
The first piece goes to zero as $\epsilon^2$ if $u \neq 0$, and its integral is $- 1/\epsilon$, so it simply gives a delta function
\begin{equation}
- \frac{16 \epsilon^2}{\pi (u^2 + 4 \epsilon^2)^2} = - \frac{1}{\epsilon} \delta(u) + {\mathcal O}(\epsilon^2),
\end{equation}
while we recognize the well-known expression of the Cauchy principal value in the second term
\begin{equation}
\lim_{\epsilon \to 0}\frac{u}{u^2 + 4 \epsilon^2} = {\cal P} \frac{1}{u}.
\end{equation}
Therefore, we find 
\begin{equation}
\frac{1}{\epsilon}\frac{d G_\epsilon(u)}{d \epsilon} =  - \frac{\delta(u)}{\epsilon} - \frac{2}{\pi} {\cal H} \frac{1}{u^2} + {\mathcal O} (\epsilon),
\end{equation}
where ${\cal H} \frac{1}{u^2} = - \frac{d}{du}  {\cal P} \frac{1}{u}$ is the derivative of the principal value, also known as the Hadamard regularization. In the following, we will denote the Hadamard regularization of an integral by a double-dashed integral 
\begin{equation}
\ddashint_{[a,b]} du \frac{f(u)}{(u-u_0)^2} \equiv \lim_{\epsilon \to 0^+} \left( \int_{a}^{u_0-\epsilon} du \frac{f(u)}{(u-u_0)^2} + \int^b_{u_0+\epsilon} du \frac{f(u)}{(u-u_0)^2} - \frac{2 f(u_0)}{\epsilon} \right),
\label{eqHadamardDef}
\end{equation}
with $x_0 \in [a,b]$.
We conclude that $g(u) = - {\cal H} \frac{1}{\pi u^2}$, so that
\begin{equation}
\frac{G_\epsilon(u)}{\epsilon^2} =  - \frac{\delta(u)}{\epsilon} - \frac{1}{\pi} {\cal H} \frac{1}{u^2} + {\mathcal O} (\epsilon).
\label{eqExpansion}
\end{equation}
This was derived explicitly for the half-plane, but conformal invariance implies this result should be true to leading order for simply connected geometries with smooth boundary.

\subsection{Explicit calculation of the boundary term in the rectangle geometry}

\begin{figure}[h]
	\begin{center}
		\begin{tikzpicture}
			\foreach \i in {-2,-1,...,4}{
				\draw (2*\i,-2.5) -- (2*\i,4.5);
				\draw (-4.5,\i) -- (8.5,\i);
			}	
			\draw[ultra thick] (0,0) rectangle (2,1);

			\draw[red, dashed] (0.2,0) -- (0.2,1);

			\foreach \i in {-1,0,...,2}{
				\foreach \j in {-1,0,...,1}{
					\node[empty dot] at (-0.2+4*\i,0.7+2*\j) {};
					\node[dot] at (0.2+4*\i,0.7+2*\j)  {};
					\node[empty dot] at (0.2+4*\i,1.3+2*\j)  {};
					\node[dot] at (-0.2+4*\i,1.3+2*\j)  {};
				}
			}

			\node[dot, red,inner sep = 1.8pt,label=right:{$(\varepsilon,y)$}] at (0.2,0.7) {};
			\node[red, empty dot] at (-0.2,0.7) {};

			\node[blue, empty dot] at (0.2,1.3) {};
			\node[dot,blue] at (-0.2,1.3) {};

			\node[blue, empty dot] at (0.2,-0.7) {};
			\node[dot,blue] at (-0.2,-0.7) {};
			\node[dot] at (0.2,-1.3) {};
			\node[empty dot] at (-0.2,-1.3) {};

		\end{tikzpicture}
	\end{center}
	\caption{The locations of the poles for the Green's function on the rectangle. The bold rectangle is the primary domain and the other rectangles are image domains. The colored poles must be regularized and also give the largest contributions in the integral. The dashed line is the integration path. Empty dots denote image charges with negative charge.}
	\label{fig:green_fcn_rect_poles}
\end{figure}
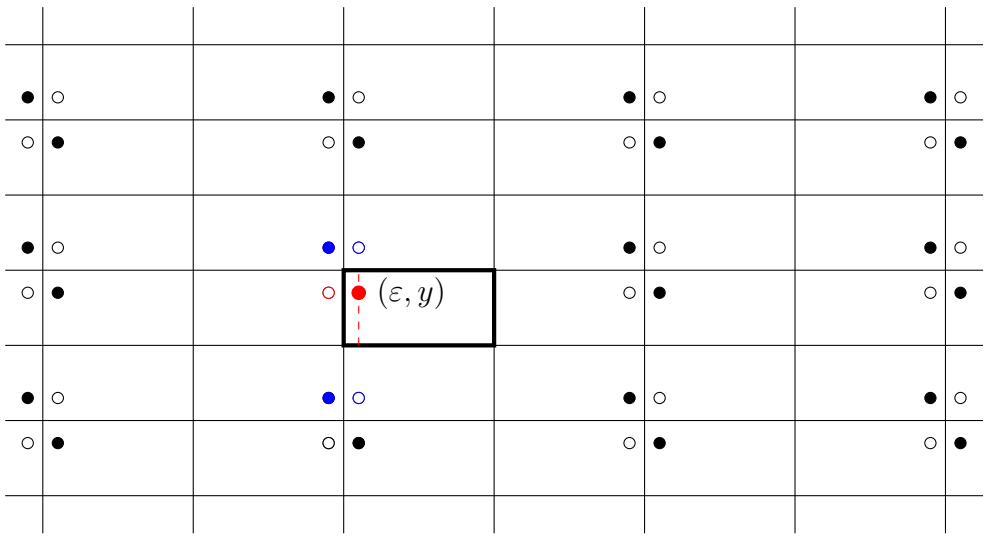

Let us now compute explicitly the value of $\gamma$ for a rectangle $L_x \times L_y$ with the entanglement cut in the $x$ direction between regions of size $\ell_x \times L_y$ and $(L_x- \ell_x) \times L_y$. Our starting point will be the formula
\begin{align}
\gamma &= \lim_{\epsilon \to 0} \frac{1}{\lambda} \int  dy\, L_\lambda(\ell_x,y) \left[ \frac{L_\lambda(\ell_x - \epsilon,y)}{\epsilon} + \frac{L_\lambda(\ell_x + \epsilon,y)}{\epsilon}  \right] \notag \\
&+ \frac{2}{\lambda} \int  dy\, dy_0\, L_\lambda(\ell_x,y) L_\lambda(\ell_x,y_0) \frac{G_\epsilon(y-y_0)}{\epsilon^2},
\label{eq:regularized_half_plane_gamma}
\end{align}
where the factor of $2$ in the second integral comes from the fact that the region $A$ and $B$ add up. The modes of the Laplacian read
\begin{equation}
L_\lambda(x,y) = \frac{2}{\sqrt{L_x L_y}} \sin \frac{\pi k_x x}{L_x}  \sin \frac{\pi k_y y}{L_y},
\end{equation}
and $\lambda= (\frac{\pi k_x}{L_x})^2+(\frac{\pi k_y}{L_y})^2$. Both terms in $\gamma$ look  badly divergent as $\epsilon \to 0$, but as explained in the main text, these divergent contributions cancel out. In principle, one should consider the full expression of the Green's function on the rectangle, but for simplicity, we will use the simplified formula~\eqref{eqGnearBdry} which should hold far away from the boundaries of the rectangle ${M}$. This is equivalent to only considering the red poles in Figure \ref{fig:green_fcn_rect_poles}. We do not expect this to change the answer significantly since the modes of the Laplacian all vanish near the corners for our geometry (the leading corrections due to these corners will be considered below). Let us expand in $\epsilon$. The first term of \ref{eq:regularized_half_plane_gamma} is easily dealt with
\begin{equation}
\frac{1}{\lambda} \int  dy\, L_\lambda(\ell_x,y) \left[ \frac{L_\lambda(\ell_x - \epsilon,y)}{\epsilon}+ \frac{L_\lambda(\ell_x + \epsilon,y)}{\epsilon}  \right]  = \frac{2}{\lambda \epsilon} \int  dy\, L^2_\lambda(\ell_x,y) + {\cal O}(\epsilon).
\end{equation}
Note that there is no ${\cal O}(1)$ term in the integral. For the second term, we will use~\eqref{eqExpansion} (which once again should hold away from the corners where the entanglement cut $\partial$ meets the boundary of ${M}$) to find
\begin{equation}
 -  \frac{2}{\lambda \epsilon} \int  dy\, L^2_\lambda(\ell_x,y) - \frac{2}{\pi \lambda} \ddashint  dy\, dy_0\, \frac{ L_\lambda(\ell_x,y) L_\lambda(\ell_x,y_0)}{(y - y_0)^2} +{\cal O}(\epsilon),
\end{equation}
We see that in the limit $\epsilon \to 0$, the divergent contributions indeed cancel out, and we get
\begin{equation}
\gamma = - \frac{2}{\pi \lambda} \ddashint  dy\, dy_0\, \frac{ L_\lambda(\ell_x,y) L_\lambda(\ell_x,y_0)}{(y - y_0)^2} .
\end{equation}
Plugging in the expression for $L_\lambda$ and changing variables to $u=y-y_0$ and $v=y_0+y$, this yields
\begin{equation}
\gamma = - \frac{4}{L_x L_y\pi \lambda} \left(\sin \frac{\pi k_x \ell_x}{L_x} \right)^2  \ddashint_{-L_y}^{L_y}  du\, \int_{\left| u \right|}^{2 L_y-\left| u \right|} dv\, \frac{\sin \frac{\pi k_y (u+v)}{2 L_y} \sin \frac{\pi k_y (v-u)}{2 L_y}}{u^2} .
\label{eqdint}
\end{equation}
The integrals can be computed explicitly using~\eqref{eqHadamardDef}, and we find 
\begin{equation}
\gamma = \frac{8}{\pi^3} \frac{(-1)^{k_y} -1+ k_y \pi  \int_{0}^{k_y \pi} dt \frac{\sin t}{t} }{k_x^2 + k_y^2 }  \left(\sin \frac{\pi k_x \ell_x}{L} \right)^2,
\label{eq:gamma_artificial_half_plane}
 \end{equation}
where we restricted ourselves to $L_x=L_y=L$ for simplicity.

This expression for the rectangle is approximate since we ignored the contribution of the corners to simplify the calculation.
Let us now evaluate how the corners modify this result. We will focus on the leading correction coming from the corner at $x=0$, $y=0$ (the blue poles in Figure \ref{fig:green_fcn_rect_poles}) so that the Green's function near the vertical boundary ($x=x_0=\epsilon$) reads
\begin{equation}
G_\epsilon(u=y-y_0,v=y+y_0) \approx \frac{1}{4 \pi} \ln \frac{u^2}{u^2 + 4 \epsilon^2}-\frac{1}{4 \pi} \ln \frac{v^2}{v^2 + 4 \epsilon^2},
\end{equation}
which vanishes for $y=0$ or $y_0=0$. The correction from the corner therefore amounts to changing $u$ to $v$ and comes with a minus sign:
\begin{equation}
\Delta \gamma_{\rm corner} =  \frac{4}{L_x \pi \lambda} \left(\sin \frac{\pi k_x \ell_x}{L_x} \right)^2  \times \frac{1}{L_y}\ddashint  du\,  dv\, \frac{\sin \frac{\pi k_y (u+v)}{2 L_y} \sin \frac{\pi k_y (v-u)}{2 L_y}}{v^2} .
\end{equation}
Evaluating the integrals for $L_x = L_y = L $, this yields
\begin{equation}
\Delta \gamma_{\rm corner} =  \frac{8}{ \pi^3}  \frac{1- (-1)^{k_y}+ k_y \pi  \int_{0}^{2 k_y \pi} dt \frac{\sin t}{t} -  k_y \pi  \int_{0}^{k_y \pi} dt \frac{\sin t}{t}}{k_x^2 + k_y^2} \left(\sin \frac{\pi k_x \ell_x}{L_x} \right)^2.
\end{equation}
The other corner at $x=0$, $y=L_y$ gives the same contribution, so that we obtain a modified expression taking into account the leading contributions of the corners as $\gamma_{\rm with \ corners} \approx \gamma + 2\Delta \gamma_{\rm corner}$, so 
\begin{equation}
\gamma_{\rm with \ corners}  =  \frac{8}{\pi^3} \frac{2 k_y \pi  \int_{0}^{2 k_y \pi} dt \frac{\sin t}{t} -(-1)^{k_y} +1- k_y \pi  \int_{0}^{k_y \pi} dt \frac{\sin t}{t} }{k_x^2 + k_y^2 }  \left(\sin \frac{\pi k_x \ell_x}{L} \right)^2.
\label{eq:gamma_with_corners}
 \end{equation}

 \subsection{Numerical Verification}

 Equation \eqref{eq:gamma_with_corners} may be verified numerically by comparing it to results on the lattice. In particular, \eqref{eq:gamma_with_corners} was compared to \eqref{eq:explicit_gamma_lattice_rectangle} for a rectangle with height $L = 9,18,\dots, 273$ for $k_x = 1$ and $k_y = 1,2,3,4$. To extract the asymptotic behavior, the numerical results were fit to
 \begin{equation}
	 \gamma(L) = a_1 + \frac{a_2}{L} + \frac{a_3}{L^2},
	 \label{eq:d_rect_fits}
 \end{equation}
 which is a good heuristic model for the results with $R^2 > 0.999$ in all cases. The coefficient $a_1$ should match the continuum result.

 \begin{center}
\begin{tabular}[h]{cllllc}
	\toprule
	\textbf{$k_y$}
	& \textbf{Analytic $\gamma$} 
	& \textbf{Fit $a_1$}
	& \textbf{Fit $a_2$}
	& \textbf{Fit $a_3$}
	& \textbf{$a_1$ Relative Error ($\%$)}\\
	\midrule
 1 & 0.05797 & 0.06478 & -0.36045 & 0.92392 & -11.7596 \\
 2 & 0.09521 & 0.09691 & -0.45105 & 0.34208 & -1.78436 \\
 3 & 0.11479 & 0.11483 & -0.58654 & 0.33662 & -0.03198 \\
 4 & 0.11931 & 0.11947 & -0.72754 & 0.96875 & -0.13306 \\
 5 & 0.1165 & 0.11653 & -0.84494 & 1.9293 & -0.02662 \\
	\bottomrule	
\end{tabular}
\end{center}
It can be seen in Figure \ref{fig:d_universal_rect} that the agreement is excellent for large $k_y$.

The main source of the error comes from the approximations made to deal with the corners. To determine this, let us consider a geometry without corners. Consider an artificial half plane: use the half-plane Green's functions but with a finite height $L$ for the boundary $\partial$. In the continuum, the answer is then given by \eqref{eq:gamma_artificial_half_plane}. On the lattice, we need the half-plane lattice Green's function, which is~\cite{Sabotka1973}
\begin{equation}
	G(x,y; x_0,y_0) = L(x-x_0,y+y_0) - L(x-x_0,y-y_0),
	\label{eq:lattice_half_plane_Green_fcn}
\end{equation}
where $L$ is a discrete version of the logarithm
\begin{equation}
	L(x,y) = \frac{1}{2\pi}\left[ \int_0^\pi \frac{1-\cos\left( y \lambda \right)\exp\left( -\n{x} \mu \right)}{\sinh \mu} \, d\lambda  - \frac{\ln 8 + 2\gamma_E}{2} \right],
\end{equation}
with $\gamma_E$ Euler's constant, $\cos \lambda + \cosh \mu = 2$ and $\mu/\lambda \to 1$ as $\lambda \to 0$. Using the same fitting procedure as before, the numerical results match the analytic ones closely, as shown in Figure \ref{fig:d_universal_hp} and the table below.
 \begin{center}
\begin{tabular}[h]{cllllc}
	\toprule
	\textbf{$k_y$}
	& \textbf{Analytic $\gamma$} 
	& \textbf{Fit $a_1$}
	& \textbf{Fit $a_2$}
	& \textbf{Fit $a_3$}
	& \textbf{$a_1$ Relative Error ($\%$)}\\
	\midrule
 1 & 0.04346 & 0.04329 & -0.41817 & 1.75342 & 0.38848 \\
 2 & 0.08621 & 0.08609 & -0.46253 & 0.79546 & 0.1478 \\
 3 & 0.1067 & 0.10663 & -0.58413 & 0.7134 & 0.06196 \\
 4 & 0.11339 & 0.11336 & -0.71384 & 1.2392 & 0.0309 \\
 5 & 0.1117 & 0.11164 & -0.82399 & 2.10472 & 0.04952 \\
 \bottomrule
\end{tabular}
\end{center}

\bibliography{EE_Excited_States_QLM_revised}

\end{document}